\documentclass[a4paper]{article}
	
\usepackage{a4wide}
\usepackage[UKenglish]{babel}
\usepackage{authblk}
\usepackage{graphicx,subfigure,caption}
\usepackage{epstopdf}
\graphicspath{{./}{figures/}}
\usepackage[colorlinks=true,linkcolor=blue,citecolor=red]{hyperref}
\usepackage{xcolor}
\usepackage{amssymb,amsthm,empheq,bbold}
\usepackage[inline]{enumitem}
\usepackage{bbm}
\usepackage{tikz}
	
\newcommand{\bA}{\mathbf{A}}
\newcommand{\R}{\mathbb{R}}
\newcommand{\N}{\mathbb{N}}
\newcommand{\bU}{\mathbf{U}}

\newcommand{\tv}{\mathrm{TV}}
\newcommand{\bVV}{\mathbf{V}}
 
\newtheorem{assumption}{Assumption}[section]
\newtheorem{theorem}[assumption]{Theorem}
\theoremstyle{remark}\newtheorem{remark}[assumption]{Remark}
\theoremstyle{definition}\newtheorem{definition}[assumption]{Definition}

\allowdisplaybreaks
		
\title{Hydrodynamic traffic flow models including random accidents: A kinetic derivation}
\author[1]{F. A. Chiarello}
\author[2]{S. G\"ottlich}
\author[2]{T. Schillinger}
\author[3]{A. Tosin}
\affil[1]{{\footnotesize University of L'Aquila, DISIM, 67100 L'Aquila, Italy}}
\affil[2]{{\footnotesize University of Mannheim, School of Business Informatics and Mathematics, 68159 Mannheim, Germany}}
\affil[3]{{\footnotesize Politecnico di Torino, Department of Mathematical Sciences ``G. L. Lagrange'', 10129 Torino, Italy}}
\date{}
	
\begin{document}
\maketitle
\begin{abstract}
We present a formal kinetic derivation of a second order macroscopic traffic model from a stochastic particle model. The macroscopic model is given by a system of hyperbolic partial differential equations (PDEs) with a discontinuous flux function, in which the traffic density and the headway are the averaged quantities. A numerical study illustrates the performance of the second order model compared to the particle approach. We also analyse numerically uncertain traffic accidents by considering statistical measures of the solution to the PDEs.
        
\medskip

        
\medskip
        
\end{abstract}
    
\section{Introduction}
Traffic flow can be modelled at different scales for example using ordinary differential equations (ODEs), kinetic equations or partial differential equations (PDEs). ODE-based models describe microscopically the behaviour of individual vehicles. In this paper, we consider especially the class of Follow-the-Leader (FTL) models \cite{gazis1961OR}, in which the vehicle dynamics are influenced by the distance to the vehicle in front.
    
The modelling of vehicular traffic can also be based on the statistical representation of interacting particle systems along the lines of the collisional kinetic theory. In this approach, vehicles are regarded as indistinguishable particles, whose pairwise interactions produce speed variations. The indistinguishability assumption allows one to describe them by means of the statistical distribution of their speed, pretty much like in the kinetic approach to gas dynamics introduced by Boltzmann. The pioneer of the kinetic approach to vehicular traffic was Prigogine~\cite{prigogine1960OR,prigogine1971BOOK}, who in the 1960s proposed to adapt the classical concepts of the statistical physics of gases to vehicles along a road, so as to obtain a mathematical representation of traffic which could serve as a link between the microscopic vehicle-wise and the macroscopic fluid dynamic descriptions. Since then, several improvements have been proposed, such as the use of Enskog-type rather than Boltzmann-type kinetic equations to capture the non-locality of vehicle interactions, see e.g.~\cite{klar1997JSP}, up to more general Povzner-Boltzmann-type equations, which are able to explain the genesis of non-local macroscopic traffic models, see e.g.~\cite{chiarello2023KRM}. The kinetic approach has proved useful also in upscaling microscopically controlled vehicle dynamics to the macroscopic scale, whereby hydrodynamic traffic models have been deduced, which incorporate consistently the effect of driver-assist or automated vehicles on the mean flow, see e.g.~\cite{chiarello2021MMS,dimarco2022JSP}.
    
Macroscopic models, which do not focus on individual vehicles but on the traffic density as an aggregate quantity, take inspiration from fluid dynamics and therefore are based on hyperbolic conservation laws. They were first introduced by Lighthill, Whitham and Richards \cite{lighthill1955PRSLA,richards1956OR} in the 1950s. Since then, first order models have been extended in several directions. Aw, Rascle and Zhang \cite{aw2000SIAP,zhang2003} introduced second order traffic models where density and speed are considered as the averaged quantities. 
  
In this paper, starting from FTL microscopic dynamics, we adopt an Enskog-type kinetic approach to obtain a second order macroscopic model in which, besides the vehicle density, the traffic flow is described by the mean \textit{headway} among the vehicles. The resulting hydrodynamic model is original in that the mean headway is treated as a second aggregate variable independent of the traffic density, whereas in classical traffic models it is empirically assumed to be proportional to the inverse of the latter. Moreover, thanks to the kinetic approach, the model is obtained as a \textit{physical limit} of fundamental particle dynamics. This introduces a remarkable difference with respect to the mainstream in the reference literature, where the link between microscopic and macroscopic descriptions of traffic is usually established by showing that selected versions of the former may be used as numerical discretisation of the latter, with convergence in appropriate particle limits. Furthermore, our model allows for an additional space dependence in the flux function for varying road capacities to model traffic accidents. For the sake of completeness, we note that hyperbolic partial differential equations with an additional space dependence in the flux have been studied before, e.g., in \cite{holden2018,karlsen2004,zhang2002TRB}, and for a first order traffic model in \cite{ThomasSimone}, where a corresponding microscopic model and its convergence are discussed.
		
As mentioned above, as an application we use our space-dependent macroscopic model to describe \textit{traffic accidents}, which we understand as capacity drops in the flux function. This idea was developed in \cite{knapp2020,ThomasSimone}. Further approaches to traffic accident modelling can be found in the literature from various disciplines: for instance, by using kinetic models \cite{freguglia2017} or by constructing Bayesian networks \cite{Mora2017,Zou2017} and recently also neural networks \cite{garcia2018, Zhao2019}. In our case, the physical limit mentioned above, along with the probabilistic/statistical setting of the kinetic theory, allows us to treat accidents as capacity drops in \textit{random}, viz. \textit{uncertain}, locations along the road, which translates into a macroscopic model featuring a realistically uncertain flux. For the sake of completeness, we report that the effect of uncertain quantities on traffic dynamics has already been taken into account in a number of other papers. Without pretending to be exhaustive, we mention that in~\cite{herty2018SIAP} uncertain lateral speeds, orthogonal to the main traffic stream, are introduced to model the displacement of vehicles across the lanes of a multi-lane road; in~\cite{herty2021SEMA-SIMAI} uncertain vehicle interactions are considered, in a homogeneous Boltzmann-type kinetic modelling framework, to explain the emergence of equilibrium speed distributions comparable with those shown by rough traffic data; in~\cite{tosin2021MCRF} a theoretical investigation, still based on concepts and tools of the collisional kinetic theory, is proposed concerning the ability of autonomous vehicles to mitigate the impact of uncertain vehicle interactions on the aggregate traffic predictions; finally, in~\cite{Brencher2020} analytical properties of conservation laws with uncertain and discontinuous flux functions are discussed.

In more detail, the plan of this work is as follows. In Section~\ref{sect:s} we introduce the underlying microscopic FTL traffic model, which we use in Section~\ref{sect:enskogDescription} to derive stochastic particle dynamics and therefrom the second order macroscopic limiting model with density and mean headway, of which we discuss some relevant analytical properties. We also show that in an appropriate regime of the parameters of the particle dynamics, the second order model relaxes towards a first order model, in which the main aggregate quantity is the traffic density while the mean headway is expressed as a function of the density derived from the interaction rules among the particles. We complete the picture by illustrating numerically the results. In Section~\ref{sect:KineticRandomAccidents} we extend the multiscale modelling framework above with the inclusion of random accidents and we undertake a computational analysis of the performances of the model with accidents using different numerical simulation strategies to capture the expected values of the traffic density and mean headway at the various scales. Finally, in Section~\ref{sect:Conclusion} we draw some conclusions and briefly sketch possible research developments.
    
\section{Microscopic dynamics with headway}
\label{sect:s}
We consider the following Follow-the-Leader (FTL) microscopic model introduced in~\cite{ThomasSimone}
\begin{equation}
    \dot{x}_i(t)=c(x_i(t))\tilde{V}\!\left(\frac{L}{x_{i+1}(t)-x_i(t)}\right), \quad i=1,\,2,\,\dots
    \label{eq:micro_model}
\end{equation}
where $L>0$ is a reference vehicle length, $x_i(t)\in\R$ the position of the $i$-th vehicle at time $t$ and $\tilde V$ a given speed function. Moreover, $c=c(x):\R\to [0,\,1]$ is a prescribed function modulating the actual speed of the vehicles depending on the \textit{road capacity} in the point $x\in\R$.

We define the distance between two consecutive vehicles $i$ and $i+1$, i.e. the \textit{headway} of the $i$-th vehicle, as
$$s_{i}(t):=x_{i+1}(t)-x_i(t),$$
whence, using~\eqref{eq:micro_model},
$$ \dot{s}_{i}(t)=c(x_{i+1}(t))\tilde{V}\!\left(\frac{L}{s_{i+1}(t)}\right)-c(x_{i}(t))\tilde{V}\!\left(\frac{L}{s_{i}(t)}\right). $$
If the vehicles $i$, $i+1$ participating in the interaction described by this equation are meant to be representative of any pair of interacting vehicles, we can drop the indices $i$, $i+1$ and define the generic positions $x:=x_i(t)$, $x_\ast:=x_{i+1}(t)$ and the pre-interaction headways $s:=s_i(t)$, $s_\ast:=s_{i+1}(t)$. Furthermore, if we assume that the variation of the headway in consequence of an interaction takes place in a small time interval of size $\gamma>0$ we can approximate $\dot{s}_i(t)\approx\frac{s_i(t+\gamma)-s_i(t)}{\gamma}$, where we identify $s':=s_i(t+\gamma)$ as the post-interaction headway. Finally, we convert the previous ODE into the following algebraic binary interaction rule
\begin{equation}
    s'=s+\gamma\bigl(c(x_\ast)V(s_\ast)-c(x)V(s)\bigr),
    \label{eq:s_interaction_rules}
\end{equation}
where we have denoted $V(s):=\tilde{V}(\frac{L}{s})$.

In addition to the FTL interaction dynamics described by~\eqref{eq:s_interaction_rules}, we consider also a spontaneous relaxation of the headway of each vehicle towards an optimal/recommended headway $H$, which we assume to be given as a function of the global density $\rho$ of traffic: $H=H(\rho)$. Hence we couple to~\eqref{eq:s_interaction_rules} a second update rule of the headway of the form
		\begin{equation}
		\label{eq:microHeadway}
		s''=s+a(H(\rho)-s),
		\end{equation}
where $a>0$ is a relaxation parameter.

On the functions $V$, $H$ we make the following assumptions:
\begin{assumption}\label{asm}
We assume that the speed $V$ is a non-negative function of the headway $s\geq 0$ with the following characteristics:
\begin{enumerate}[label=(\roman*)]
\item it is differentiable and monotonically increasing:
$$ V'(s)>0, \quad \forall\,s\in\R_+; $$
\item there exists a constant $C>0$ such that
$$ 0\leq V(s)\leq C s, \quad \forall\,s\in\R_+. $$
\end{enumerate}
Furthermore, we assume that the optimal/recommended headway $H$ is a non-negative, differentiable and monotonically decreasing function of the traffic density $\rho\geq 0$.
\end{assumption}
         
Assumption~\ref{asm}(i) is quite natural from the modelling point of view: the larger the distance between two consecutive vehicles the faster they travel. Analogously, the assumption on the monotonicity of $H$ has a modelling value, because it implies that the more congested the traffic the closer the vehicles are forced to stay.

Conversely, Assumption~\ref{asm}(ii) is needed in order to guarantee the physical consistency of the binary interaction rule~\eqref{eq:s_interaction_rules}, in particular the fact that $s'\geq 0$ for all $s,\,s_\ast\geq 0$. Using Assumption~\ref{asm}(ii), we easily check that this condition is satisfied if $\gamma\leq C$. Finally, as far as the update rule~\eqref{eq:microHeadway} is concerned, we notice that the analogous condition of physical consistency $s''\geq 0$ for all $s\geq 0$ is guaranteed if $a\leq 1$, thanks to the non-negativity of $H$.
   		
\section{Enskog-type kinetic description and hydrodynamics}
\label{sect:enskogDescription}
We consider the superposition of the FTL interaction dynamics~\eqref{eq:s_interaction_rules} and the Optimal Headway (OH) relaxation dynamics~\eqref{eq:microHeadway} and we show that, in the hydrodynamic limit, two types of macroscopic models can be obtained. If OH dynamics happen at a much slower rate than FTL dynamics then we get an inhomogeneous second-order macroscopic model featuring the traffic density and the mean headway as hydrodynamic parameters. Conversely, OH and FTL dynamics happen at comparable rates then we get a first order Lighthill-Whitham-Richards type-model featuring only the traffic density as hydrodynamic parameter.

To obtain these results, we rely on an Enskog-type collisional kinetic description of the system of interacting particles subject to the rules~\eqref{eq:s_interaction_rules},~\eqref{eq:microHeadway}.

\subsection{Slow relaxation regime}
\label{sect:slow_relax}
We consider a large ensemble of indistinguishable vehicles, each of which is identified by the dimensionless position $X_t\in\R$ and dimensionless headway $S_t\in\R_+$ at time $t>0$. Motivated by the rules~\eqref{eq:s_interaction_rules},~\eqref{eq:microHeadway}, we consider the following discrete-in-time stochastic particle model:
		\begin{equation}
		\begin{cases}
		X_{t+\Delta{t}}=X_t+c\left(X_t\right)V(S_t)\Delta{t}, \qquad\\
		S_{t+\Delta{t}}=S_t+\gamma\Theta\Bigl(c\left(X^\ast_t\right)V(S^\ast_{t})-c\left(X_t\right)V(S_t)\Bigr)+\Xi a(H(\rho)-S_t),
		\end{cases}
		\label{eq:particle.random}
		\end{equation}
		where $\Delta{t}>0$ is a (small) time step. Moreover, $\Theta,\,\Xi\in\{0,\,1\}$ are Bernoulli random variables describing whether during the time step $\Delta{t}$ a randomly chosen vehicle with microscopic state $(X_t,S_t)$:
  \begin{enumerate*}[label=(\roman*)]
  \item updates ($\Theta=1$) or not ($\Theta=0$) its headway owing to an FTL interaction with the leading vehicle in $X^\ast_t$;
  \item updates ($\Xi=1$) or not ($\Xi=0$) its headway because of an OH relaxation towards the optimal headway $H(\rho)$.
  \end{enumerate*}
  In more detail, we let
		\begin{equation}
		\Theta\sim\operatorname{Bernoulli}(\Delta{t}), \qquad \Xi\sim\operatorname{Bernoulli}(\varepsilon\Delta{t}),
		\label{eq:Bernoulli.slow_OH}
		\end{equation}
		thereby assuming that the probability for either updates to happen is proportional to $\Delta{t}$. The parameter $\varepsilon>0$ is used to differentiate the rate of OH relaxation from that of FTL interactions. In particular, here we assume that OH relaxation is much slower than FTL interactions, i.e.
		$$ \varepsilon\ll 1. $$
		Furthermore, we need $\Delta{t}\leq 1$ for consistency.
  
To reach an aggregate statistical description of our particle system we introduce the kinetic distribution function $f=f(x,s,t):\R\times\R_ +\times (0,\,+\infty)\to\R_+$ of the microscopic state $(x,s)$ of a generic representative vehicle at time $t$. In essence, $f(x,s,t)\,dx\,ds$ gives the probability that a vehicle has a position comprised between $x$ and $x+dx$ and a headway comprised between $s$ and $s+ds$ at time $t$. Then, by standard arguments, see e.g.,~\cite{pareschi2013BOOK}, averaging~\eqref{eq:particle.random} and taking the continuous-time limit $\Delta{t}\to 0^+$ we obtain formally that $f$ satisfies the following equation:
\begin{align}
    \begin{aligned}[b]
        \partial_t\int_0^{+\infty}\varphi(s)f(x,s,t)\,ds &+ \partial_x\left(c(x)\int_0^{+\infty}\varphi(s)V(s)f(x,s,t)\,ds\right) \\
        &=\frac{1}{2}\int_0^{+\infty}\int_0^{+\infty}(\varphi(s')-\varphi(s))f(x,s,t)f(x+\eta,s_\ast,t)\,ds\,ds_\ast\\
        &\phantom{=}+\varepsilon\int_0^{+\infty}(\varphi(s'')-\varphi(s))f(x,s,t)\,ds,
    \end{aligned}
    \label{eq:kinetic}
\end{align}
for every choice of $\varphi:\R_+\to\R$, which here plays the role of a test function, where $s'$ and $s''$ on the right-hand side are given by~\eqref{eq:s_interaction_rules},~\eqref{eq:microHeadway}, respectively.

Notice that~\eqref{eq:kinetic} is the weak form of a collisional kinetic equation in which we have assumed that vehicles interact when they are at a distance $\eta>0$ from each other (cf. the first term on the right-hand side). In other words, with respect to the notation used in the interaction rule~\eqref{eq:s_interaction_rules}, we have assumed $x_\ast=x+\eta$. Therefore,~\eqref{eq:kinetic} is an \textit{Enskog-type} kinetic equation, which, as discussed in~\cite{klar1997JSP}, is more appropriate than a Boltzmann-type equation to model vehicular traffic. The main reason is that a non-locality in space of the interactions is necessary in order to reproduce   density waves possibly travelling backwards in spite of the non-negativity of the microscopic car speeds.
		
In order to make~\eqref{eq:kinetic} more amenable to further analytical investigations, we assume that the non-locality $\eta$ is sufficiently small so that we can approximate
\begin{equation}
    f(x+\eta, s_\ast, t)\approx f(x,s_\ast,t)+\eta \partial_x f(x,s_\ast,t).
    \label{eq:f.approx}
\end{equation}
An analogous approximation applies also to the term
\begin{equation}
    c(x_\ast)=c(x+\eta)\approx c(x)+c'(x)\eta,
    \label{eq:c.approx}
\end{equation}
contained in $s'$, cf.~\eqref{eq:s_interaction_rules}, whence, assuming $\varphi$ smooth,
\begin{equation*}
\begin{aligned}
    \varphi(s') &= \varphi\Bigl(s+\gamma\bigl(c(x_\ast)V(s_\ast)-c(x)V(s)\bigr)\Bigr)\\
    &= \varphi\bigl(\tilde{s}'+\gamma\eta c'(x)V(s_\ast)\bigr)\\
    &\approx\varphi(\tilde{s}')+\varphi'(\tilde{s}')\gamma\eta c'(x)V(s_\ast),
\end{aligned}
\end{equation*}
where we have set
\begin{equation}
    \tilde{s}':=s+\gamma c(x)\bigl(V(s_\ast)-V(s)\bigr).
    \label{eq:tilde_s'}
\end{equation}
Plugging these approximations into~\eqref{eq:kinetic} and enforcing the equality we get the following approximated kinetic equation:
\begin{align}
\begin{aligned}[b]
    \partial_t\int_0^{+\infty}\varphi(s)f(x,s,t)\,ds &+ \partial_x\left(c(x)\int_0^{+\infty}\varphi(s)V(s)f(x,s,t)\,ds\right) \\
    &=\frac{1}{2}\int_0^{+\infty}\int_0^{+\infty}(\varphi(\tilde{s}')-\varphi(s))f(x,s,t)f(x,s_\ast,t)\,ds\,ds_\ast \\
    &\phantom{=}+\frac{\gamma\eta}{2}c'(x)\int_0^{+\infty}\int_0^{+\infty}\varphi'(\tilde{s}')V(s_\ast)f(x,s,t)f(x,s_\ast,t)\,ds\,ds_\ast \\
    &\phantom{=}+\frac{\eta}{2}\int_0^{+\infty}\int_0^{+\infty}(\varphi(\tilde{s}')-\varphi(s))f(x,s,t)\partial_x f(x,s_\ast,t) \,ds\,ds_\ast \\
    &\phantom{=}+\frac{\gamma\eta^2}{2}c'(x)\int_0^{+\infty}\int_0^{+\infty}\varphi'(\tilde{s}')f(x,s,t)\partial_x f(x,s_\ast,t) \,ds\,ds_\ast \\
    &\phantom{=}+\varepsilon\int_0^{+\infty}(\varphi(s'')-\varphi(s))f(x,s,t)\,ds.
\end{aligned}
\label{eq:Enskog_approx}
\end{align}

\subsubsection{Hydrodynamic limit}
To pass from the kinetic description~\eqref{eq:Enskog_approx} to the hydrodynamic regime, we use the parameter $\varepsilon$ as a sort of \textit{Knudsen number}. Specifically, we scale time and space as
\begin{equation}
t\to\frac{t}{\varepsilon}, \qquad x\to\frac{x}{\varepsilon},
\label{eq:hyp_scal3}
\end{equation}
whence $\partial_t\to\varepsilon\partial_t$, $\partial_x\to\varepsilon\partial_x$, $c'(x)\to \varepsilon c'(x)$, and consequently we rewrite~\eqref{eq:Enskog_approx} as
\begin{align}
\begin{aligned}[b]
    \partial_t\int_0^{+\infty}\varphi(s)f^\varepsilon(x,s,t)\,ds &+ \partial_x\left(c(x)\int_0^{+\infty}\varphi(s)V(s)f^\varepsilon(x,s,t)\,ds\right) \\
    &=\frac{1}{2\varepsilon}\int_0^{+\infty}\int_0^{+\infty}(\varphi(\tilde{s}'))-\varphi(s))f^\varepsilon(x,s,t)f^\varepsilon(x,s_\ast,t)\,ds\,ds_\ast \\
    &\phantom{=}+\frac{\gamma\eta}{2}c'(x)\int_0^{+\infty}\int_0^{+\infty}\varphi'(\tilde{s}')V(s_\ast)f^\varepsilon(x,s,t)f^\varepsilon(x,s_\ast,t)\,ds\,ds_\ast \\
    &\phantom{=}+\frac{\eta}{2}\int_0^{+\infty}\int_0^{+\infty}(\varphi(\tilde{s}')-\varphi(s))f^\varepsilon(x,s,t)\partial_xf^\varepsilon(x,s_\ast,t)\,ds\,ds_\ast \\
    &\phantom{=}+\frac{\varepsilon\gamma\eta^2}{2}c'(x)\int_0^{+\infty}\int_0^{+\infty}\varphi'(\tilde{s}')V(s_\ast)f^\varepsilon(x,s,t)\partial_xf^\varepsilon(x,s_\ast,t)\,ds\,ds_\ast \\
    &\phantom{=}+\int_0^{+\infty}(\varphi(s'')-\varphi(s))f^\varepsilon(x,s,t)\,ds,
\end{aligned}
\label{eq:kinetic_scaled}
\end{align}
where $f^\varepsilon(x,s,t):=f(\frac{x}{\varepsilon},s,\frac{t}{\varepsilon})$ denotes the distribution function parameterised by $\varepsilon$. Next, we introduce the following definition.
\begin{definition} \label{def:collinv}
We call \textit{collisional invariant} of the kinetic equation~\eqref{eq:kinetic_scaled} any quantity $\varphi:\R_+\to\R$ such that
$$ \int_0^{+\infty}\int_0^{+\infty}(\varphi(\tilde{s}')-\varphi(s))f^\varepsilon(x,s,t)f^\varepsilon(x,s_\ast,t)\,ds\,ds_\ast=0. $$
\end{definition}
It is not difficult to check that $\varphi(s)=1$ and $\varphi(s)=s$ are collisional invariants in the sense of Definition~\ref{def:collinv}. Plugging these two quantities into~\eqref{eq:kinetic_scaled} we obtain therefore, for every $\varepsilon>0$,
\begin{align}
    \left\{
        \begin{aligned}[c]
            \partial_t\int_0^{+\infty}f^\varepsilon(x,s,t)\,ds &+ \partial_x\left(c(x)\int_0^{+\infty}V(s)f^\varepsilon(x,s,t)\,ds\right)=0, \\
            \partial_t\int_0^{+\infty}sf^\varepsilon(x,s,t)\,ds &+ \partial_x\left(c(x)\int_0^{+\infty}sV(s)f^\varepsilon(x,s,t)\,ds\right) \\
            &=\frac{\gamma\eta}{2}c'(x)\int_0^{+\infty}\int_0^{+\infty}V(s_\ast)f^\varepsilon(x,s,t)f^\varepsilon(x,s_\ast,t)\,ds\,ds_\ast \\
            &\phantom{=}+\frac{\gamma\eta}{2}c(x)\int_0^{+\infty}\int_0^{+\infty}(V(s_\ast)-V(s))f^\varepsilon(x,s,t)\partial_xf^\varepsilon(x,s_\ast,t)\,ds\,ds_\ast \\
            &\phantom{=}+\frac{\varepsilon\gamma\eta^2}{2}c'(x)\int_0^{+\infty}\int_0^{+\infty}V(s_\ast)f^\varepsilon(x,s,t)\partial_xf^\varepsilon(x,s_\ast,t)\,ds\,ds_\ast \\
            &\phantom{=}+a\int_0^{+\infty}(H(\rho)-s)f^\varepsilon(x,s,t)\,ds.
        \end{aligned}
    \right.
    \label{eq:kinetic_hydro}
\end{align}

On the other hand, in the \textit{hydrodynamic limit} $\varepsilon\to 0^+$ equation~\eqref{eq:kinetic_scaled} implies that the limit distribution $f^0$ satisfies formally
\begin{equation}
    \int_0^{+\infty}\int_0^{+\infty}(\varphi(\tilde{s}')-\varphi(s))f^0(x,s,t)f^0(x,s_\ast,t)\,ds\,ds_\ast=0,
    \label{eq:local_Maxwellian}
\end{equation}
for \textit{every} observable quantity $\varphi$. We call a solution $f^0$ to this equation a \textit{local Maxwellian}, i.e. an equilibrium distribution of the headway $s$ for fixed $x$, $t$.

Assume we are given a local Maxwellian. Then, passing to the hydrodynamic limit also in~\eqref{eq:kinetic_hydro} we discover:
\begin{align}
    \left\{
        \begin{aligned}[c]
            \partial_t\int_0^{+\infty}f^0(x,s,t)\,ds &+ \partial_x\left(c(x)\int_0^{+\infty}V(s)f^0(x,s,t)\,ds\right)=0, \\
            \partial_t\int_0^{+\infty}sf^0(x,s,t)\,ds &+ \partial_x\left(c(x)\int_0^{+\infty}sV(s)f^0(x,s,t)\,ds\right) \\
            &=\frac{\gamma\eta}{2}c'(x)\int_0^{+\infty}\int_0^{+\infty}V(s_\ast)f^0(x,s,t)f^0(x,s_\ast,t)\,ds\,ds_\ast \\
            &\phantom{=}+\frac{\gamma\eta}{2}c(x)\int_0^{+\infty}\int_0^{+\infty}(V(s_\ast)-V(s))f^0(x,s,t)\partial_xf^0(x,s_\ast,t)\,ds\,ds_\ast \\
            &\phantom{=}+a\int_0^{+\infty}(H(\rho)-s)f^0(x,s,t)\,ds.
        \end{aligned}
    \right.
    \label{eq:kinetic_Maxwellian}
\end{align}

Clearly, any local Maxwellian is defined up to the collisional invariants. This means, in particular, that $f^0$ is parameterised by:
\begin{itemize}
\item the traffic density
$$ \rho(x,t):=\int_0^{+\infty}f^0(x,s,t)\,ds; $$
\item the mean headway
$$ h(x,t):=\frac{1}{\rho(x,t)}\int_0^{+\infty}sf^0(x,s,t)\,ds. $$
\end{itemize}
Indeed equation~\eqref{eq:local_Maxwellian}, which for $\varphi(s)=1,\,s$ is trivially satisfied because of Definition~\ref{def:collinv}, cannot determine univocally the zeroth and first order $s$-moments of $f^0$. Consequently, if an explicit expression of $f^0$ is available, system~\eqref{eq:kinetic_Maxwellian} may provide the evolution equations for $\rho$, $h$, thus the macroscopic counterpart of the particle traffic model~\eqref{eq:particle.random}.

It is not difficult to check, by direct substitution in~\eqref{eq:local_Maxwellian}, that the \textit{monokinetic} distribution
\begin{equation}
    f^0(x,s,t)=\rho(x,t)\delta(s-h(x,t)),
    \label{eq:maxwellian_h}
\end{equation}
$\delta(\cdot)$ being the Dirac distribution, is a local Maxwellian. Uniqueness of~\eqref{eq:maxwellian_h} is hard to obtain for a completely general speed function $V$. Nevertheless, it can be recovered, under additional assumptions on $V$, in the \textit{quasi-invariant regime} of the particle dynamics~\eqref{eq:particle.random}, i.e. a regime reminiscent of the \textit{grazing collision regime} of the classical kinetic theory in which every particle interaction produces a little change of microscopic state but interactions are quite frequent. See~\cite{MMS2021} for further details.

Using~\eqref{eq:maxwellian_h} in~\eqref{eq:kinetic_Maxwellian} we get, after some computations, the \textit{second order} macroscopic model
\begin{equation}
    \begin{cases}
        \partial_t\rho+\partial_x\bigl(c(x)V(h)\rho\bigr)=0, \\[3mm]
	\partial_t(\rho h)+\partial_x\bigl(c(x)V(h)\rho h\bigr)=\dfrac{\gamma}{2}\rho^2\eta\partial_x\bigl(c(x)V(h)\bigr)+a\rho(H(\rho)-h),
    \end{cases}
    \label{eq:macroheadway}
\end{equation}
in the hydrodynamic parameters $\rho$, $h$.

\subsubsection{Analytical insights into the macroscopic system}
Despite the general derivation of a macroscopic description incorporating both FTL and relaxation dynamics, here we focus on system~\eqref{eq:macroheadway} without relaxation term:
\begin{equation}
    \begin{cases}
        \partial_t\rho+\partial_x\bigl(c(x)V(h)\rho\bigr)=0, \\[3mm]
	\partial_t(\rho h)+\partial_x\bigl(c(x)V(h)\rho h\bigr)=\dfrac{\gamma}{2}\rho^2\eta\partial_x\bigl(c(x)V(h)\bigr),
    \end{cases}
    \label{eq:macroheadway_withoutrelax}
\end{equation}
with $(x,t)\in \R\times [0,\,+\infty)$. System~\eqref{eq:macroheadway_withoutrelax} constitutes the hydrodynamic counterpart of the original microscopic FTL model~\eqref{eq:micro_model}.

We notice that we can write~\eqref{eq:macroheadway_withoutrelax} in conservative form defining the pressure function $p(\rho)=\frac{\gamma}{2}\eta\rho$, i.e.
\begin{equation}
    \begin{cases}
	\partial_t\rho+\partial_x\bigl(c(x)V(h)\rho\bigr)=0, \\[2mm]
	\partial_t\bigl(\rho(h+p(\rho))\bigr)+\partial_x\bigl(c(x)V(h)\rho(h+p(\rho))\bigr)=0.
    \end{cases}
    \label{eq:macroheadway_conservative1}
\end{equation}
		
Now we rewrite system~\eqref{eq:macroheadway_conservative1} as 
\begin{equation}
    \begin{cases}
        \partial_t\rho+\partial_x\bigl(cV(h)\rho\bigr)=0, \\[2mm]
	\partial_t\bigl(\rho(h+p(\rho))\bigr)+\partial_x\bigl(cV(h)\rho(h+p(\rho))\bigr)=0, \\[2mm]
	\partial_tc=0.
    \end{cases}
    \label{eq:macroheadway_conservative_3}
\end{equation}
This is a $3\times3$ system for $\bVV:=(\rho,\,\rho (h+p(\rho)),\,c)^T$
		%\end{remark}
		and we complement it with the initial datum  
		\begin{equation}\label{eq:initial_cond}
		\bar u(x)=(\bar \rho, \bar h, \bar c)(x) \quad \text{s.t.} \quad \bar \rho(x)>0, \,\bar h(x)>0,\, \bar c(x)>0.
		\end{equation}
		
	The quasilinear vector form of the	system~\eqref{eq:macroheadway_conservative_3} is
		$$ \partial_t\bU+\bA(\bU)\partial_x\bU=\mathbf{0},$$
		with $\bU:=(\rho,\,h,c)^T$ and
		$$	\bA(\bU):=
		\begin{pmatrix}
		c V(h) & V'(h)c \rho & V(h)\rho\\
		0 & c \left(V(h)-\frac{\gamma}{2}\eta\rho V'(h)\right)& -\frac{\gamma}{2}\eta\rho V(h)\\
		0 & 0 & 0
		\end{pmatrix}.
		$$ The eigenvalues $\lambda_1,\,\lambda_2,\,\lambda_3, $ and eigenvectors $r_1,\,r_2, \,r_3,$ of this matrix are 
		$$ \lambda_3=c V\!\left(h\right) \quad \text{with} \quad r_3=(1,\,0,\, 0),$$
		
		$$ \lambda_2=c \left(V\!\left(h\right)-\frac{\gamma}{2}\eta \rho V'\!\left(h\right)\right)
		\quad \text{with} \quad r_2=\left(1,\,-\frac{\gamma}{2}\eta,\,0\right),$$
		and
		$$ \lambda_1=0
		\quad \text{with} \quad r_1=\left(V(h)\rho,\,-\frac{\gamma}{2}\eta V(h)\rho,\, c \left(\rho \frac{\gamma}{2}\eta V'(h)-V(h) \right)\right). $$
		Since the eigenvalues are real and $\bA(\bU)$ is diagonalisable, system~\eqref{eq:macroheadway} is hyperbolic. Furthermore, under the Assumptions $V'(h)>0,\,V(h)>0$ and $V(h)> \frac{\gamma}{2}\eta\rho V'(h)$, it results $\lambda_1<\lambda_2<\lambda_3$, then the system is strictly hyperbolic. No characteristic speed is greater than the flow speed. Hence~\eqref{eq:macroheadway_withoutrelax} complies with the Aw-Rascle consistency condition.
		The first characteristic field and the third characteristic field are linearly degenerate: $\nabla\lambda_1\cdot r_1=0,\, \nabla\lambda_3\cdot r_3=0$, thus the associated waves are contact discontinuities. Conversely, the second characteristic field is genuinely nonlinear: $\nabla\lambda_2\cdot r_2\neq 0$, hence the associated waves are either shocks or rarefactions, if the velocity function is such that $V'(h)-\frac{\rho}{2}V''(h)\neq 0$. It is worth noticing that, choosing $V(h):=\frac{h}{1+h}$ we get $\nabla\lambda_2\cdot r_2<0 .$
		In this setting, in order to prove the global existence of entropy solutions, we can apply~\cite[Theorem 7.1]{Bressan} under a quite restrictive assumption on the total variation of the initial datum.
		\begin{theorem}Let us consider the set $\Omega$ such that for every $(\rho,h,c)\in \Omega$ the following conditions are verified: 
			\begin{itemize}
				\item $V(h),\, c$ and $V'(h)$ are strictly greater than zero;
				\item $V(h)- \frac{\gamma}{2}\eta\rho V'(h)\neq 0$;
				\item $V'(h)-\frac{\rho}{2}V''(h)\neq 0$.
			\end{itemize}
			For every compact $K \subset \Omega$ there exists a constant $\delta>0$ with the following property. For every initial condition $\bar u$ with 
			\begin{equation}
			\tv{(\bar u})\leq \delta, \quad \lim_{x\to -\infty} \bar u(x)\in K,
			\end{equation}
			the Cauchy problem~\eqref{eq:macroheadway_conservative_3}-\eqref{eq:initial_cond} has a weak entropy solution $u(t,x)=(\rho, h, c)(t,x),$ defined for all $t\geq 0.$ If a convex entropy $\tilde \eta$ is given, then the Cauchy problem~\eqref{eq:macroheadway_conservative_3}-\eqref{eq:initial_cond} admits an $\tilde\eta-$admissible solution.
		\end{theorem}
In order to give more analytical insights,  let us consider the Riemann problem for our system~\eqref{eq:macroheadway_conservative_3}, with initial data
	\begin{equation*}
	\bar u(0,x):=\begin{cases}
	(\rho^-,h^-,c^-), \quad x<0,\\
	(\rho^+,h^+,c^+), \quad x>0,\\
	\end{cases}
	\end{equation*}
	assuming $\rho^\pm,\,h^\pm,\,c^\pm>0.$ 
	By $r_1, r_2, r_3$ the rarefaction curves through $(\rho^-,h^-,c^-)$ are obtained solving the following Cauchy problems.
  From $r_1$ we get 
	\begin{align*}
	\begin{cases}
	\dot \rho&= V(h)\rho,\\
	\dot h&=-\frac{\gamma}{2}\eta V(h)\rho,\\
	\dot c&=c\left(\rho \frac{\gamma}{2}\eta V'(h)-V(h)\right).
	\end{cases}
	\end{align*}
	This yields to
	\begin{align*}
	\begin{cases}
	\dot \rho&= V(h)\rho,\\
	\dot h&=-\frac{\gamma}{2}\eta \dot \rho,\\
	\dot c&=c\left(\rho \frac{\gamma}{2}\eta V'(h)-V(h)\right),
	\end{cases}
	\end{align*}
	then, we end up with the implicit curve 
	\begin{align*}
	&\mathbf{R_1}:
	\sigma_1 \to \left(\rho^- e^{\int_0^{\sigma_1} V(h(\sigma_1))d\sigma_1}, h^- -\frac{\gamma}{2}\eta (\rho(\sigma_1)-\rho^-), c^-e^{\int_0^{\sigma_1} \left(-V(h(\sigma_1))+\frac{\gamma}{2}\eta \rho V'(h(\sigma_1))\right)d\sigma_1}\right). 
	\end{align*}
	
	From $r_2$ we write 
	\begin{align*}
	\begin{cases}
	\frac{dh}{d\rho}&=-\frac{\gamma}{2}\eta,\\
	h(\rho^-)&=h^-,\\
	\dot c&=0.
	\end{cases}
	\end{align*}
	This yields to the curve 
		\begin{equation*}
	\mathbf{R_2}: \sigma_2 \to (\sigma_2+\rho^-,h^--\frac{\gamma}{2}\eta \sigma_2,c^-),
	\end{equation*}
	that can be rewritten as
	\begin{equation*}
	\mathbf{R_2}=\left\{(\rho,h,c):\:h-h^-=-\frac{\gamma}{2}\eta\left(\rho-\rho^- \right),\, c=c^-\right\}.
	\end{equation*}

 From $r_3$ we have 
	\begin{align*}
	\begin{cases}
	\dot \rho &= 1,\\
	\dot h &= 0,\\
	\dot c &= 0.
	\end{cases}
	\end{align*}
	This yields the curve 
	\begin{equation*}
	\mathbf{R_3}: \sigma_3 \to (\sigma_3+\rho^-,h^-,c^-).
	\end{equation*}

    The shock curves $\mathbf{S_1},\,\mathbf{S_2}$ and $\mathbf{S_3}$ through $(\rho^-, h^-, c^-)$ are derived from the Rankine-Hugoniot conditions
	\begin{align*}
	\lambda (\rho^- -\rho)&=c^-V(h^-)\rho^- -c V(h)\rho,\\
	\lambda\left(\rho^-(h^-+p(\rho^-))-\rho (h+p(\rho))\right)&=c^- \rho^-V(h^-)\left(h^-+p(\rho^-)\right)-c \rho V(h)\left(h+p(\rho)\right),\\
	\lambda (c^--c)&=0.
	\end{align*}
	We can observe by a straightforward computation that $\mathbf{S_1}$ coincides with the rarefaction curve $\mathbf{R_1}\,$, $\mathbf{S_2}$ coincides with the rarefaction curve $\mathbf{R_2}$ and $\mathbf{S_3}$ coincides with $\mathbf{R_3},$ being the characteristic fields associated to $r_1$ and $r_3$ linearly degenerate. 
	\begin{remark}
     It is interesting to notice that system~\eqref{eq:macroheadway_conservative_3} reduces to the following $2\times 2$ system when $c$ is constant:
\begin{equation*}
    \begin{cases}
        \partial_t\rho+c\partial_x\bigl(V(h)\rho\bigr)=0, \\[3mm]
	\partial_t\bigl(\rho(h+p(\rho))\bigr)+c\partial_x\bigl(V(h)\rho(h+p(\rho))\bigr)=0.
    \end{cases}
    \label{eq:macroheadway_conservative}
\end{equation*}
This system of conservation laws belongs to the Temple class, see~\cite{DafermosGeng, Temple}, and deserves more analytical attention, for this reason a deeper study will be done in a future work.
\end{remark}
		
\subsection{Fast relaxation regime}
We now  analyse the case in which the FTL and the relaxation dynamics take place at the same rate. This means that, in place of~\eqref{eq:Bernoulli.slow_OH}, we consider
$$ \Theta,\,\Xi\sim\operatorname{Bernoulli}(\Delta{t}), $$
which produces the following weak form of the Enskog-type kinetic equation:
\begin{align}
    \begin{aligned}[b]
	\partial_t\int_0^{+\infty}\varphi(s)f(x,s,t)\,ds &+ \partial_x\left(c(x)\int_0^{+\infty}\varphi(s)V(s)f(x,s,t)\,ds\right) \\
	&= \frac{1}{2}\int_0^{+\infty}\int_0^{+\infty}(\varphi(s')-\varphi(s))f(x,s,t)f(x+\eta,s_\ast,t)\,ds\,ds_\ast \\
	&\phantom{=} +\int_0^{+\infty}(\varphi(s'')-\varphi(s))f(x,s,t)\,ds.
    \end{aligned}
    \label{eq:kinetic_fast}
\end{align}
in place of~\eqref{eq:kinetic}. The difference with respect to the latter is that the two terms at the right-hand side are now of the same order of magnitude. Repeating the approximations~\eqref{eq:f.approx},~\eqref{eq:c.approx} under the assumption of small $\eta$ yields
\begin{align}
\begin{aligned}[b]
    \partial_t\int_0^{+\infty}\varphi(s)f(x,s,t)\,ds &+ \partial_x\left(c(x)\int_0^{+\infty}\varphi(s)V(s)f(x,s,t)\,ds\right) \\
    &=\frac{1}{2}\int_0^{+\infty}\int_0^{+\infty}(\varphi(\tilde{s}')-\varphi(s))f(x,s,t)f(x,s_\ast,t)\,ds\,ds_\ast \\
    &\phantom{=}+\frac{\gamma\eta}{2}c'(x)\int_0^{+\infty}\int_0^{+\infty}\varphi'(\tilde{s}')V(s_\ast)f(x,s,t)f(x,s_\ast,t)\,ds\,ds_\ast \\
    &\phantom{=}+\frac{\eta}{2}\int_0^{+\infty}\int_0^{+\infty}(\varphi(\tilde{s}')-\varphi(s))f(x,s,t)\partial_x f(x,s_\ast,t)\,ds\,ds_\ast \\
    &\phantom{=}+\frac{\gamma\eta^2}{2}c'(x)\int_0^{+\infty}\int_0^{+\infty}\varphi'(\tilde{s}')f(x,s,t)\partial_x f(x,s_\ast,t)\,ds\,ds_\ast \\
    &\phantom{=}+\int_0^{+\infty}(\varphi(s'')-\varphi(s))f(x,s,t)\,ds,
\end{aligned}
\label{eq:Enskog_approx-fast}
\end{align}
in place of~\eqref{eq:Enskog_approx}, $\tilde{s}'$ being again given by~\eqref{eq:tilde_s'}.

\subsubsection{Hydrodynamic limit}
Under the scaling~\eqref{eq:hyp_scal3}, the kinetic equation~\eqref{eq:Enskog_approx-fast} in the scaled distribution function $f^\varepsilon$ becomes:
\begin{align}
\begin{aligned}[b]
    \partial_t\int_0^{+\infty}\varphi(s)f^\varepsilon(x,s,t)\,ds &+ \partial_x\left(c(x)\int_0^{+\infty}\varphi(s)V(s)f^\varepsilon(x,s,t)\,ds\right) \\
    &=\frac{1}{2\varepsilon}\int_0^{+\infty}\int_0^{+\infty}(\varphi(\tilde{s}')-\varphi(s))f^\varepsilon(x,s,t)f^\varepsilon(x,s_\ast,t)\,ds\,ds_\ast \\
    &\phantom{=}+\frac{\gamma\eta}{2}c'(x)\int_0^{+\infty}\int_0^{+\infty}\varphi'(\tilde{s}')V(s_\ast)f^\varepsilon(x,s,t)f^\varepsilon(x,s_\ast,t)\,ds\,ds_\ast \\
    &\phantom{=}+\frac{\eta}{2}\int_0^{+\infty}\int_0^{+\infty}(\varphi(\tilde{s}')-\varphi(s))f^\varepsilon(x,s,t)\partial_xf^\varepsilon(x,s_\ast,t)\,ds\,ds_\ast \\
    &\phantom{=}+\frac{\varepsilon\gamma\eta^2}{2}c'(x)\int_0^{+\infty}\int_0^{+\infty}\varphi'(\tilde{s}')f^\varepsilon(x,s,t)\partial_xf^\varepsilon(x,s_\ast,t)\,ds\,ds_\ast \\
    &\phantom{=}+\frac{1}{\varepsilon}\int_0^{+\infty}(\varphi(s'')-\varphi(s))f^\varepsilon(x,s,t)\,ds.
\end{aligned}
\label{eq:Enskog_approx-fast.scaled}
\end{align}

At this point, we introduce the following new definition of collisional invariants, which for~\eqref{eq:Enskog_approx-fast.scaled} replaces Definition~\ref{def:collinv}:
\begin{definition}
\label{def:collinv.2}
We call \textit{collisional invariant} of the kinetic equation~\eqref{eq:Enskog_approx-fast.scaled} any quantity $\varphi:\R_+\to\R$ such that
$$ \frac{1}{2}\int_0^{+\infty}\int_0^{+\infty}(\varphi(\tilde{s}')-\varphi(s))f^\varepsilon(x,s,t)f^\varepsilon(x,s_\ast,t)\,ds\,ds_\ast
    +\int_0^{+\infty}(\varphi(s'')-\varphi(s))f^\varepsilon(x,s,t)\,ds=0. $$
\end{definition}
It is immediate to check that $\varphi(s)=1$ is again a collisional invariant whereas $\varphi(s)=s$ is not. Therefore, plugging $\varphi(s)=1$ into~\eqref{eq:Enskog_approx-fast.scaled} we are left, for every $\varepsilon>0$, with the equation
$$ \partial_t\int_0^{+\infty}f^\varepsilon(x,s,t)\,ds+\partial_x\left(c(x)\int_0^{+\infty}V(s)f^\varepsilon(x,s,t)\,ds\right)=0, $$
which, passing to the hydrodynamic limit $\varepsilon\to 0^+$, yields formally
\begin{equation}
    \partial_t\int_0^{+\infty}f^0(x,s,t)\,ds+\partial_x\left(c(x)\int_0^{+\infty}V(s)f^0(x,s,t)\,ds\right)=0,
    \label{eq:kinetic_Maxwellian.2}
\end{equation}
$f^0$ being a local Maxwellian which satisfies
$$ \frac{1}{2}\int_0^{+\infty}\int_0^{+\infty}(\varphi(\tilde{s}')-\varphi(s))f^0(x,s,t)f^0(x,s_\ast,t)\,ds\,ds_\ast
    +\int_0^{+\infty}(\varphi(s'')-\varphi(s))f^0(x,s,t)\,ds=0, $$
for every observable quantity $\varphi:\R_+\to\R$. For $\varphi(s)=s$ this relationship gives
$$ \rho(H(\rho)-h)=0, $$
$\rho$, $h$ being the density and the mean of $f^0$. Next, by direct substitution we discover that the monokinetic distribution
$$ f^0(x,s,t)=\rho(x,t)\delta\bigl(s-H(\rho(x,t))\bigr), $$
parameterised by $\rho$ (i.e., the macroscopic parameter associated with the only collisional invariant) is a local Maxwellian. Again, uniqueness of such a local Maxwellian is hard to obtain for a generic speed function $V$ but can be guaranteed at least in the quasi-invariant regime under additional assumptions on $V$, cf.~\cite{MMS2021}.

Inserting the $f^0$ above into~\eqref{eq:kinetic_Maxwellian.2} we finally get the \textit{first order} macroscopic model
\begin{equation}
    \partial_t\rho+\partial_x\bigl(c(x)\rho V(H(\rho))\bigr)=0,
    \label{eq:macroFirstOrder}
\end{equation}
in the hydrodynamic parameter $\rho$. Notice that with $H(\rho)=1/\rho$ this becomes the equation considered in~\cite{ThomasSimone}.

\subsection{Numerical simulations}
\label{sect:numericalSimulaitons}
\subsubsection{Discretisations}
\label{sect:Discretizations}
Consider an equispaced spatial grid $(\tilde{x}_i)_{i=1,\dots,\tilde{N}}$ with step size $\Delta x>0$ and a temporal grid $(t_j)_{j = 1,\dots,M}$ and step size $\Delta t>0$.
		The particle model \eqref{eq:particle.random} can directly be solved using a sufficiently small step size of $\Delta t \leq \min \lbrace 1, \frac{1}{\varepsilon} \rbrace$.
		The microscopic model \eqref{eq:micro_model} is solved using the explicit Euler scheme, i.e.
		\begin{align*}
		    x_{i}^{j+1} &= x_i^j + \Delta t c(x_i^j)\tilde{V}\left(\frac{L}{x_{i+1}^j-x_i^j}, \right),
		\end{align*}
		with stepsize $\Delta t \leq \frac{1}{\|c\|_\infty \|\tilde{V}\|_\infty}$ and $x_i^j$ denotes the position of the $i$-th vehicle at time $t_j$. The velocity function is set to $\tilde{V}(x)=1-x$.
		
		We apply the Lax-Friedrichs scheme to investigate the numerical behaviour of the proposed macroscopic models \eqref{eq:macroheadway} and \eqref{eq:macroFirstOrder}. Even though the scheme has diffusive properties it rebuilds the main properties of the model, especially for small stepsizes. For the second order macroscopic model \eqref{eq:macroheadway} we define $z_i^j = \rho_i^j h_i^j$ and use a splitting approach for the second equation. First, the advection step is performed, then the diffusion part is taken into account.
		\begin{align*}
		\begin{cases}
		\rho_i^{j+1} & = \frac{\rho_{i-1}^j + \rho_{i+1}^j}{2} - \frac{\Delta t}{2 \Delta x} \left(c(x_{i+1})V(h_{i+1}^j)\rho_{i+1}^j - c(x_{i-1})V(h_{i-1}^j)\rho_{i-1}^j   \right),\\
		\tilde{z}_i^{j+1} &= \frac{\rho_{i-1}^j h_{i-1}^j + \rho_{i+1}^j h_{i+1}^j}{2} - \frac{\Delta t}{2 \Delta x} \left(c(x_{i+1})V(h_{i+1}^j)\rho_{i+1}^j h_{i+1}^j - c(x_{i-1})V(h_{i-1}^j)\rho_{i-1}^j h_{i-1}^j   \right),\\
		z_i^{j+1} &= \tilde{z}_i^j + \Delta t \left(\frac{\gamma}{2} (\rho_i^j)^2\eta \frac{c(x_{i+1})V(h_{i+1}^j) - c(x_i)V(h_i^j)}{\Delta x} + a \rho_i^j (H(\rho_i^j) - h_i^j) \right),\\
		h_i^{j+1} &= \frac{z_i^{j+1}}{\rho_i^{j+1}}.
		\end{cases}
		\end{align*}
		The discretization of the first order model \eqref{eq:macroFirstOrder} is similarly given by
			\begin{align*}
		\rho_i^{j+1} & = \frac{\rho_{i-1}^j + \rho_{i+1}^j}{2} - \frac{\Delta t}{2 \Delta x} \left(c(x_{i+1})V(h_{i+1}^j)\rho_{i+1}^j - c(x_{i-1})V(h_{i-1}^j)\rho_{i-1}^j   \right).
		\end{align*}
		For stability of the numerical scheme the CFL-condition is ensured for both the first and second order model
		\begin{align*}
		    \frac{\Delta t}{\Delta x} \|c\|_\infty \|V\|_\infty\leq 1.
		\end{align*}

	%	Let $z_i^j$ be the considered quantity the scheme reads
	%	\begin{align*}
	%	z_i^{j+1} = \frac{z_{i-1}^j + z_{i+1}^j}{2} - \frac{\Delta t}{2\Delta x} \left(f(z_{i+1}^j) - f(z_{i-1}^j) \right).
	%	\end{align*}
	%	For the equations with non-vanishing right-hand side we use a splitting algorithm and in every time step first perform the Lax-Friedrichs iteration and then apply the diffusion component.
	
\subsubsection{Particle-macro and micro-macro comparison}
		We consider a road given by the interval $[-4,4]$ with periodic boundary conditions and set the initial conditions 
		\begin{equation*}
		\rho_0(x):=\begin{cases}0.15 \quad &x<0\\
		0.1 \quad &x\geq0,\\
		\end{cases} \quad  h_0(x):=\begin{cases} 0.8 \quad &x<0\\
		0.95 \quad &x\geq0.\\
		\end{cases}
		\end{equation*}
		Let $\delta>0$, then we set
		\begin{align*}
		c_\text{road}(x) = \begin{cases}
		1 & x \in [-4,-2-\delta) \cup [2+\delta, 4]\\
		0.6 & x \in [-2+\delta, 2- \delta]\\
		-\frac{0.2}{\delta}x + 0.8-\frac{0.4}{\delta} & x \in [-2-\delta, -2+\delta]\\
		\frac{0.2}{\delta}x + 0.8-\frac{0.4}{\delta} & x \in [2-\delta, 2+\delta].
		\end{cases}
		\end{align*}
		In the following we set $\delta = \frac{1}{10}$. The road capacity is reduced on the area of $[-2,2]$ which for example may be caused by an accident. For the microscopic simulation $c_\text{road}$ needs to be sufficiently smooth. Therefore, we add a linear interpolation in an $\delta$-range around the discontinuities. 
		In the numerical tests we consider the following functions
		\begin{equation}
		\label{eq:def_velo_headway}
		V(h):=\frac{h}{h+1} \quad \text{and}\quad H(\rho):=\frac{1}{1+\rho}.
		\end{equation}
		
\begin{figure}[!t]
    \centering
    \includegraphics[width=\textwidth]{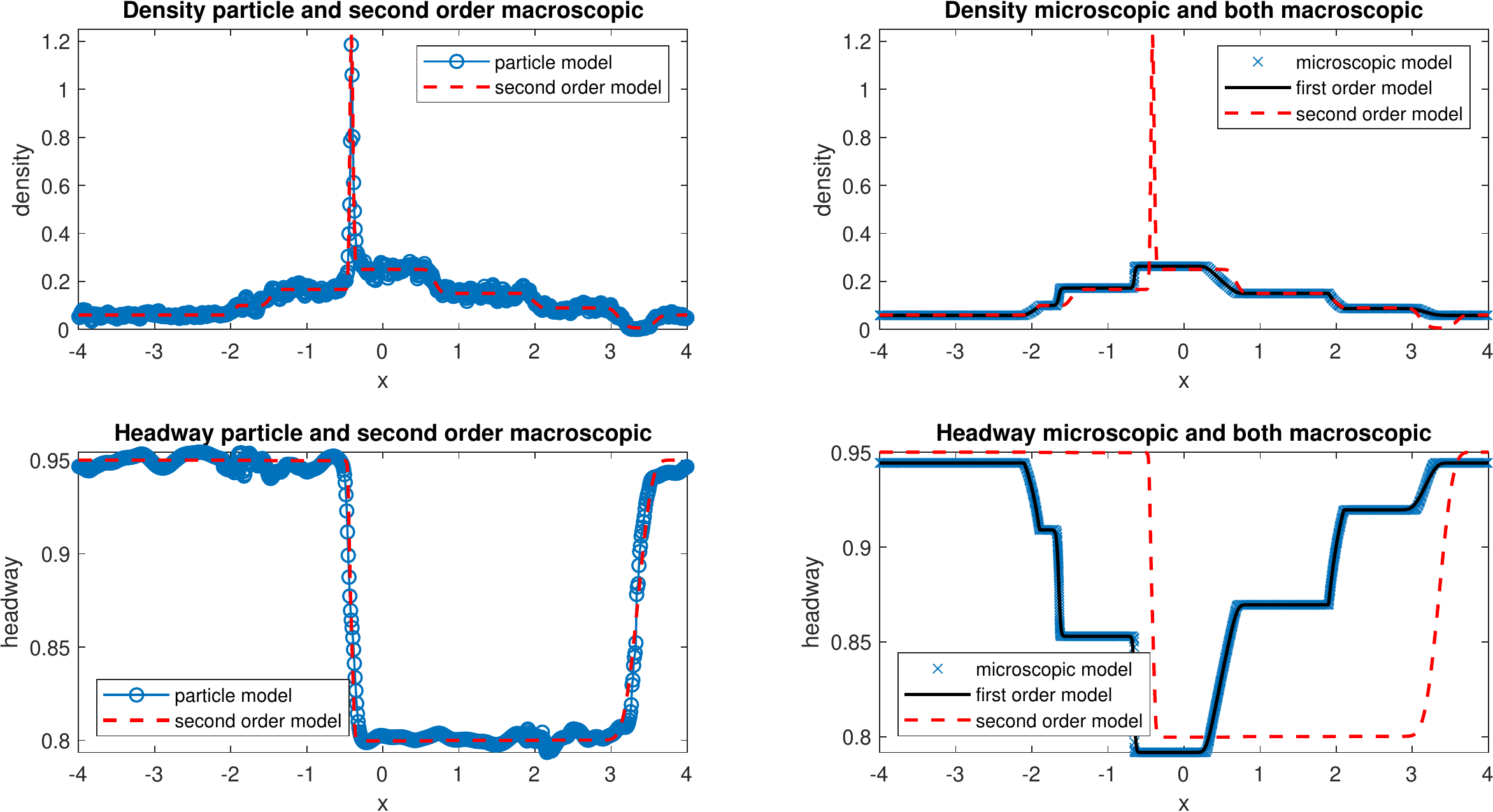}
    \caption{Comparison of density and headway between the particle model and the second order macroscopic model (left) and the microscopic model together with first and second order macroscopic model (right) without relaxation ($a=0$).}
    \label{fig:comparison_a=0}
\end{figure}

\begin{figure}[!t]
    \centering
    \includegraphics[width=\textwidth]{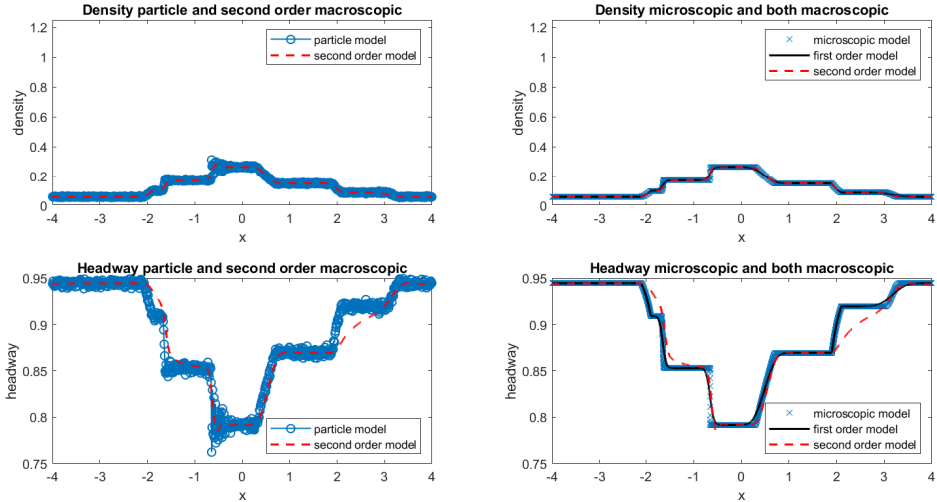}
    \caption{Comparison of density and headway between the particle model and the second order macroscopic model (left) and the microscopic model together with first and second order macroscopic model (right) with relaxation parameter $a=1$.}
    \label{fig:comparison_a=0.5}
\end{figure}

		We consider a time horizon of $T = 10$. We set $\Delta x = 10^{-2}$, $\varepsilon = 10^{-3}$, $\Delta t = \varepsilon$ and $10^6$ particles in the particle simulation and $\Delta x = 2\cdot10^{-4}$,$\Delta t = 2\cdot 10^{-4}$, $\gamma = 0.5$, $\eta=10^{-2}$ in the macroscopic simulations to reduce the diffusion of the Lax-Friedrichs-Scheme. In the microscopic model we consider $N=10^4$ vehicles and a vehicle length of $L=\frac{1}{N}$. The initial vehicle positions are arranged such that the local densities correspond to $\rho_0$. To visualize the effects of the relaxation parameter $a$ of the second order macroscopic model \eqref{eq:macroheadway} we show simulations for $a=0$ (Figure \ref{fig:comparison_a=0}) and $a=1$ (Figure \ref{fig:comparison_a=0.5}). Only in the particle and the second order macroscopic model, the headway is an state variable in the system. To be able also to compare headways from the microscopic and first order macroscopic model, in these models we compute the headways artificially according to $H(\rho)$ defined in \eqref{eq:def_velo_headway}.
	

  %\begin{figure}[ht!]
%		    \centering
%		    \includegraphics[scale=.55]{./figures/a_1v2_stepsize2_10000.eps}
%		    \caption{Comparison of density and headway between the particle model and the second order macroscopic model (left) and the microscopic model together with first and second order macroscopic model (right) with relaxation parameter $a=1$.}
%		    \label{fig:comparison_a=0.5}
%		\end{figure}
		In the left half of Figure \ref{fig:comparison_a=0} we observe a good match of the particle model and the second order macroscopic model for both, the density and the headway. On the right half for the comparison of the microscopic model and the first and second order macroscopic model, the microscopic and the first order macroscopic model show the same behaviour for on both scales. In the density, the second order model is still close, but shows significant deviations in for example the increase around $x=-0.5$ which has been captured by the particle simulation but not the microscopic or first order macroscopic model. The difference is driven by the different evolution of the headway in the second order model. Comparing Figure \ref{fig:comparison_a=0} and \ref{fig:comparison_a=0.5} we observe that increasing the relaxation parameter $a$ pushes the particle and second order macroscopic model closer to microscopic model and first order macroscopic model. For both the second order macroscopic model and the particle model, the peak in the density around $x=-0.5$ is reduced. A similar behaviour is observable in the headway illustration. Compared to $a=0$, in Figure \ref{fig:comparison_a=0.5} the headways of the second order macroscopic model and especially the particle model very strongly tend to the ones from the microscopic model. %This results in numerical evidence that for a high relaxation parameter the second order macroscopic model is similar to the microscopic and first order macroscopic model. %Furthermore in the second order macroscopic model, we observe an oscillation around $x=-0.5$, which can also be found in the particle model, but on a significantly smaller scale.

\section{Kinetic and macroscopic descriptions including random accidents}
\label{sect:KineticRandomAccidents}
		
\begin{figure}[!t]
    \centering
    \begin{tikzpicture}[scale=0.8]
\draw[->] (-5.5,0) -- (5,0);
\draw[->] (-0.25,-0.3) -- (-0.25,5);
\draw[dashed] (-3,2) -- (-3,3);
\draw[dashed] (2.5,2) -- (2.5,3);
\draw[dotted] (-3,0) -- (-3,2);
\draw[dotted] (2.5,0) -- (2.5,2);
\draw (2.5,-0.3) node {$Y$};
\draw (-3,-0.3) node {$-Y$};
\draw[line width=1.5,-] (-3,2) -- (2.5,2);
%\draw (0.1,1.5) node {$0.6$};
\draw[line width=1.5,-] (-5,3) -- (-3,3);
\draw[line width=1.5,-] (2.5,3) -- (4.5,3);
\draw (0,3) node {$1$};
\draw (5.3,0) node {$x$};
\draw (0.6,5) node {$c(x;Y)$};
\draw (0,-0.3) node {$0$};
\end{tikzpicture}
    \caption{Prototypical road capacity function parameterised by the random extent of an accident. Outside the uncertain stretch $[-Y,\,Y]$ we have $c=1$, whereas within the uncertain stretch $[-Y,\,Y]$ we have $c<1$.}
    \label{fig:random_size}
\end{figure}
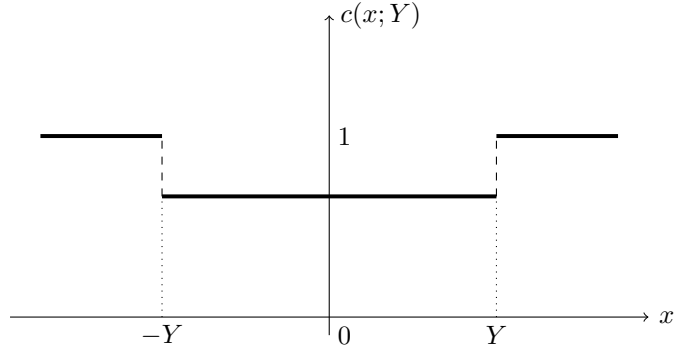

In this section, we go back to the particle model~\eqref{eq:particle.random} \textit{without} relaxation ($\Xi\equiv 0$) and we include \textit{random accidents} understood as a reduction of the road capacity in \textit{uncertain} locations. In more detail, we consider:
\begin{equation}
    \begin{cases}
        X_{t+\Delta{t}}=X_t+c(X_t;Y)V(S_t)\Delta{t}, \qquad\\
	S_{t+\Delta{t}}=S_t+\gamma\Theta\Bigl(c(X^\ast_t;Y)V(S^\ast_{t})-c(X_t;Y)V(S_t)\Bigr),
    \end{cases}
    \label{eq:particle.random_2}
\end{equation}
where $Y\in\R_+$ is a bounded random variable parameterising the road capacity function $c$, such that $[-Y,\,Y]$ is the uncertain interval within which an accident taking place at $x=0$ affects the traffic flow by reducing the road capacity, see Figure~\ref{fig:random_size}. Thus, $2Y$ is the uncertain size of the accident.
\begin{remark}
Contrary to~\eqref{eq:particle.random}, here we disregard the relaxation term in the particle model~\eqref{eq:particle.random_2} because for this application we do not intend to compare first and second order macroscopic dynamics. Therefore, we stick to the original motivating FTL model~\eqref{eq:micro_model}.
\end{remark}
		
\subsection{Enskog-type kinetic description and its hydrodynamic limit}
The (weak) Enskog-type description of the particle dynamics~\eqref{eq:particle.random_2} is formally analogous to~\eqref{eq:kinetic} up to dropping the second integral at the right-hand side:
\begin{align}
    \begin{aligned}[b]
        \partial_t\int_0^{+\infty}\varphi(s)f(x,s,t;Y)\,ds &+ \partial_x\int_0^{+\infty}\varphi(s)c(x;Y)V(s)f(x,s,t;Y)\,ds \\
	&=\frac{1}{2}\int_0^{+\infty}\int_0^{+\infty}(\varphi(s')-\varphi(s))f(x,s,t;Y)f(x+\eta,s_\ast,t;Y)\,ds\,ds_\ast.
    \end{aligned}
    \label{eq:kinetic_random}
\end{align}
However, now also the interaction rule~\eqref{eq:s_interaction_rules} is parameterised by the random variable $Y$, which plays indeed the role of a random parameter in the whole equation. Therefore, in~\eqref{eq:kinetic_random} it results
$$ s'=s+\gamma\bigl(c(x_\ast;Y)V(s_\ast)-c(x;Y)V(s)\bigr). $$
On the whole, the solution $f$ to~\eqref{eq:kinetic_random} depends on $Y$ as a parameter, thus we write $f=f(\cdot,\cdot,\cdot;Y)$.

If the non-locality $\eta$ of the interactions is sufficiently small, we may approximate~\eqref{eq:kinetic_random} by (cf.~\eqref{eq:Enskog_approx}):
\begin{align*}
    \partial_t\int_0^{+\infty}\varphi(s)f(x,s,t;Y)\,ds &+ \partial_x\left(c(x;Y)\int_0^{+\infty}\varphi(s)V(s)f(x,s,t;Y)\,ds\right) \\
    &=\frac{1}{2}\int_0^{+\infty}\int_0^{+\infty}(\varphi(\tilde{s}')-\varphi(s))f(x,s,t;Y)f(x,s_\ast,t;Y)\,ds\,ds_\ast \\
    &\phantom{=}+\frac{\gamma\eta}{2}\partial_xc(x;Y)\int_0^{+\infty}\int_0^{+\infty}\varphi'(\tilde{s}')V(s_\ast)f(x,s,t;Y)f(x,s_\ast,t;Y)\,ds\,ds_\ast \\
    &\phantom{=}+\frac{\eta}{2}\int_0^{+\infty}\int_0^{+\infty}(\varphi(\tilde{s}')-\varphi(s))f(x,s,t;Y)\partial_xf(x,s_\ast,t;Y)\,ds\,ds_\ast \\
    &\phantom{=}+\frac{\gamma\eta^2}{2}\partial_xc(x;Y)\int_0^{+\infty}\int_0^{+\infty}\varphi'(\tilde{s}')f(x,s,t)\partial_xf(x,s_\ast,t;Y)\,ds\,ds_\ast,
\end{align*}
where
$$ \tilde{s}':=s+\gamma c(x;Y)\bigl(V(s_\ast)-V(s)\bigr). $$
Under the scaling~\eqref{eq:hyp_scal3} this yields
\begin{align*}
    \partial_t\int_0^{+\infty}\varphi(s)f^\varepsilon(x,s,t;Y)\,ds &+ \partial_x\left(c(x;Y)\int_0^{+\infty}\varphi(s)V(s)f^\varepsilon(x,s,t;Y)\,ds\right) \\
    &=\frac{1}{2\varepsilon}\int_0^{+\infty}\int_0^{+\infty}(\varphi(\tilde{s}')-\varphi(s))f^\varepsilon(x,s,t;Y)f^\varepsilon(x,s_\ast,t;Y)\,ds\,ds_\ast \\
    &\phantom{=}+\frac{\gamma\eta}{2}\partial_xc(x;Y)\int_0^{+\infty}\int_0^{+\infty}\varphi'(\tilde{s}')V(s_\ast)f^\varepsilon(x,s,t;Y)f^\varepsilon(x,s_\ast,t;Y)\,ds\,ds_\ast \\
    &\phantom{=}+\frac{\eta}{2}\int_0^{+\infty}\int_0^{+\infty}(\varphi(\tilde{s}')-\varphi(s))f^\varepsilon(x,s,t;Y)\partial_xf^\varepsilon(x,s_\ast,t;Y)\,ds\,ds_\ast \\
    &\phantom{=}+\frac{\varepsilon\gamma\eta^2}{2}\partial_xc(x;Y)\int_0^{+\infty}\int_0^{+\infty}\varphi'(\tilde{s}')f^\varepsilon(x,s,t)\partial_xf^\varepsilon(x,s_\ast,t;Y)\,ds\,ds_\ast.
\end{align*}

Based on Definition~\ref{def:collinv}, that we may reapply in this case, we see that $\varphi(s)=1,\,s$ are collisional invariants for this kinetic equation. Using them we get, for every $\varepsilon>0$, the system of equations
\begin{align*}
    \left\{
        \resizebox{\textwidth}{!}{$
        \begin{aligned}[c]
            \partial_t\int_0^{+\infty}f^\varepsilon(x,s,t;Y)\,ds &+ \partial_x\left(c(x;Y)\int_0^{+\infty}V(s)f^\varepsilon(x,s,t;Y)\,ds\right)=0, \\
            \partial_t\int_0^{+\infty}sf^\varepsilon(x,s,t;Y)\,ds &+ \partial_x\left(c(x;Y)\int_0^{+\infty}sV(s)f^\varepsilon(x,s,t;Y)\,ds\right) \\
            &=\frac{\gamma\eta}{2}\partial_xc(x;Y)\int_0^{+\infty}\int_0^{+\infty}V(s_\ast)f^\varepsilon(x,s,t;Y)f^\varepsilon(x,s_\ast,t;Y)\,ds\,ds_\ast \\
            &\phantom{=}+\frac{\gamma\eta}{2}c(x;Y)\int_0^{+\infty}\int_0^{+\infty}(V(s_\ast)-V(s))f^\varepsilon(x,s,t;Y)\partial_xf^\varepsilon(x,s_\ast,t;Y)\,ds\,ds_\ast \\
            &\phantom{=}+\frac{\varepsilon\gamma\eta^2}{2}\partial_xc(x;Y)\int_0^{+\infty}\int_0^{+\infty}V(s_\ast)f^\varepsilon(x,s,t;Y)\partial_xf^\varepsilon(x,s_\ast,t;Y)\,ds\,ds_\ast,
        \end{aligned}
        $}
    \right.
\end{align*}
which, in the hydrodynamic limit $\varepsilon\to 0^+$, converges formally to
\begin{align}
    \left\{
        \resizebox{.91\textwidth}{!}{$
        \begin{aligned}[c]
            \partial_t\int_0^{+\infty}f^0(x,s,t;Y)\,ds &+ \partial_x\left(c(x;Y)\int_0^{+\infty}V(s)f^0(x,s,t;Y)\,ds\right)=0, \\
            \partial_t\int_0^{+\infty}sf^0(x,s,t;Y)\,ds &+ \partial_x\left(c(x;Y)\int_0^{+\infty}sV(s)f^0(x,s,t;Y)\,ds\right) \\
            &=\frac{\gamma\eta}{2}\partial_xc(x;Y)\int_0^{+\infty}\int_0^{+\infty}V(s_\ast)f^0(x,s,t;Y)f^0(x,s_\ast,t;Y)\,ds\,ds_\ast \\
            &\phantom{=}+\frac{\gamma\eta}{2}c(x;Y)\int_0^{+\infty}\int_0^{+\infty}(V(s_\ast)-V(s))f^0(x,s,t;Y)\partial_xf^0(x,s_\ast,t;Y)\,ds\,ds_\ast,
        \end{aligned}
        $}
    \right.
    \label{eq:system_f0_uncertain}
\end{align}
$f^0$ being the local Maxwellian, which satisfies
$$ \int_0^{+\infty}\int_0^{+\infty}(\varphi(\tilde{s}')-\varphi(s))f^0(x,s,t;Y)f^0(x,s_\ast,t;Y)\,ds\,ds_\ast=0, $$
for every observable $\varphi$.

Similarly to Section~\ref{sect:slow_relax}, the monokinetic distribution
$$ f^0(x,s,t;Y)=\rho(x,t;Y)\delta(s-h(x,t;Y)), $$
turns out to be a local Maxwellian, whose uniqueness can be established e.g., in the quasi-invariant regime along the lines of the theory developed in~\cite{chiarello2021MMS}. Notice that the uncertainty brought by the random variable $Y$ naturally translates on the hydrodynamic parameters $\rho$, $h$ associated with the two collisional invariants above. With such an $f^0$,~\eqref{eq:system_f0_uncertain} specialises as
\begin{equation}
    \begin{cases}
        \partial_t\rho+\partial_x\bigl(c(x;Y)V(h)\rho\bigr)=0, \\[3mm]
        \partial_t(\rho h)+\partial_x\bigl(c(x;Y)V(h)\rho h\bigr)=\dfrac{\gamma}{2}\rho^2\eta\partial_x\bigl(c(x;Y)V(h)\bigr),
    \end{cases}
    \label{eq:random_macro2}
\end{equation}
which, as a matter of fact, coincides with the second order macroscopic model without relaxation~\eqref{eq:macroheadway_withoutrelax} but, in this case, with an uncertain solution $(\rho,\,h)$ due to the uncertain extent of the accident parameterising the road capacity function $c$.

In the classical spirit of uncertainty quantification, the family of uncertain solutions $\{(\rho,\,h)\}_Y$ to~\eqref{eq:random_macro2} can be post-processed to average the uncertainty out to a deterministic macroscopic description. This can be done by computing the expectations of $\rho$, $h$ with respect to the law of $Y$. In more detail, assume that the latter is expressed by a probability distribution $g=g(y)$, then the following mean density and headway can be defined:
\begin{equation}
    \mathbb{E}_Y[\rho(x,t;Y)]:=\int_\R\rho(x,t;y)g(y)\,dy, \qquad \mathbb{E}_Y[h(x,t;Y)]:=\int_\R h(x,t;y)g(y)\,dy.
    \label{eq:expectationMacro}
\end{equation}
Notice that both $\mathbb{E}_Y[\rho(x,t;Y)]$, $\mathbb{E}_Y[h(x,t;Y)]$ are functions of $x$, $t$ but they are not, in general, a solution to either~\eqref{eq:random_macro2} or any other specific macroscopic model.

\subsection{Microscopic and macroscopic numerical simulations}
\label{sect:SimulationsAccidents}
We consider the microscopic model from (\ref{eq:micro_model}) in which the capacity function $c$ additionally depends on the accident size random variable $Y$ introduced in Section \ref{sect:KineticRandomAccidents}
\begin{equation}
    \dot x_i(t;Y)=c(x_i(t;Y);Y) \tilde V\left(\frac{L}{x_{i+1}(t;Y)-x_i(t;Y)}\right), \quad i=1,2,...
    \label{eq:micro_model_Y}
\end{equation}
where
\begin{align}
    \label{eq:caRandom}
    c(x;Y) = 1 - 0.4\cdot\mathbbm{1}_{[-Y,\,Y]}(x).
\end{align}
{The accident size $Y$ is set to $Y=2Z +1$, where $Z$ is a Beta distributed random variable with parameters $\alpha,\beta>0$ taking values on $[0,1]$, i.e., the probability density function of $Z$ for $x\in[0,1]$ is given by
\begin{align*}
    \varphi_Y(x,\alpha,\beta) = \frac{\Gamma(\alpha+\beta)}{\Gamma(\alpha) + \Gamma(\beta)} x^{\alpha-1}(1-x)^{\beta-1}.
\end{align*} Note that choosing $\alpha=\beta=1$ results in the special case of a uniform distribution on the interval $[1,3]$ for $Y$.}
%The accident size $Y \sim \mathcal{U}([1,3])$ is a uniformly distributed random variable on $[1,3]$. 
This construction corresponds to an accident centered at $x=0$ with size $2Y$. In this chapter we are interesting in the evolution of the local densities
\begin{align}
    \rho_i^{(N)}(t;Y)=\frac{L}{x^{(N)}_{i+1}(t;Y)-x^{(N)}_i(t;Y)}, \qquad i=1,\,\dots,\,N-1,
    \label{eq: locDensities}
\end{align} 
dependent on the realization of the random variable $Y$. Especially, we consider the piecewise constant function 
\begin{align}
\label{eq: locDensities2}
\rho^{(N)}(x,t)=\rho_i^{(N)}(t), \qquad x\in [x_i(t),\,x_{i+1}(t))
\end{align}
which will be used to compare to the evolution of the macroscopic densities. Parameters are chosen as in Section \ref{sect:Discretizations}. 

%To be able to compare the realizations for different accident sizes we define a underlying grid that coincides with the macroscopic one and has a step size of $\Delta x = \frac{1}{1000}$. For any point of the grid, we then assign the local density of the closest vehicle. Without the introduction of such a grid, in different realizations vehicles may be located at different positions, such that it is not clear how to compare the densities.\\
		
In a second step, we consider the second order macroscopic model from~\eqref{eq:random_macro2} and set the accident capacity function as in~\eqref{eq:caRandom}. We are interested in the quantities $\mathbb{E}_Y[\rho(x,T;Y)]$ and $\mathbb{E}_Y[\rho^{(N)}(x,T;Y)]$, where $\rho(x,T;Y), \rho^{(N)}(x,T;Y)$ are random variables of the densities depending on the realization of $Y$.
		
A Monte Carlo simulation of $2 \cdot 10^3$ samples is used to approximate the expectation of the densities at each point of the spatial grid. We choose the same parameters for the microscopic and macroscopic model as in Section \ref{sect:numericalSimulaitons}, except for the temporal and spatial step sizes of the Lax-Friedrichs-scheme we set $\Delta x = \Delta t = 10^{-3}$ to reduce the computational effort.

\begin{figure}[!t]
\begin{minipage}{0.49\textwidth}
    \centering
    \includegraphics[width=\textwidth]{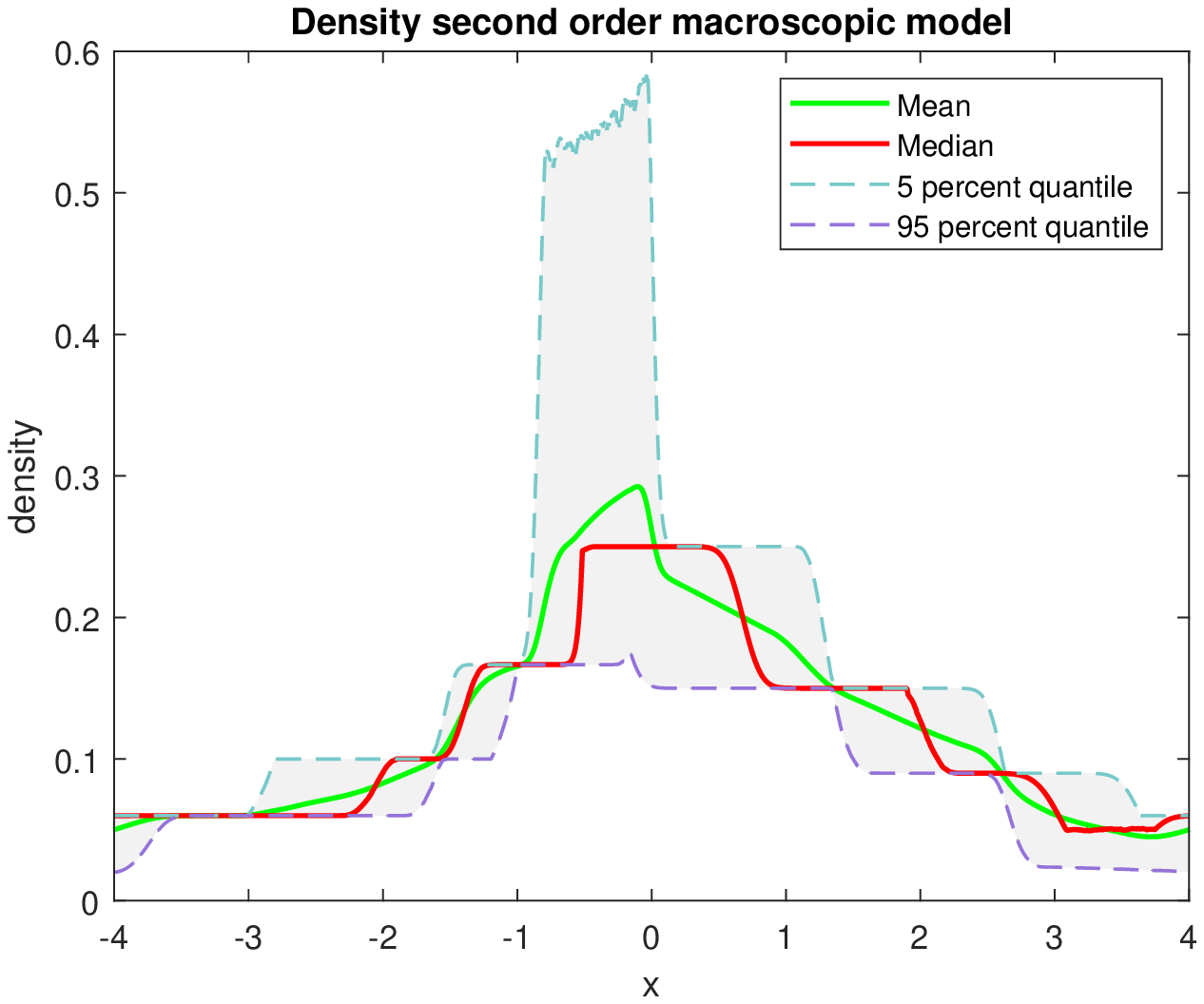}
    \captionof{figure}{Mean, median and 90 percent confidence interval of the density evolution of the second order macroscopic model \eqref{eq:random_macro2} at $T=10$ for uniformly distributed $Y$.}
    \label{fig:confidenceMacro}
\end{minipage}~
\begin{minipage}{0.49\textwidth}
    \centering
    \includegraphics[width=\textwidth]{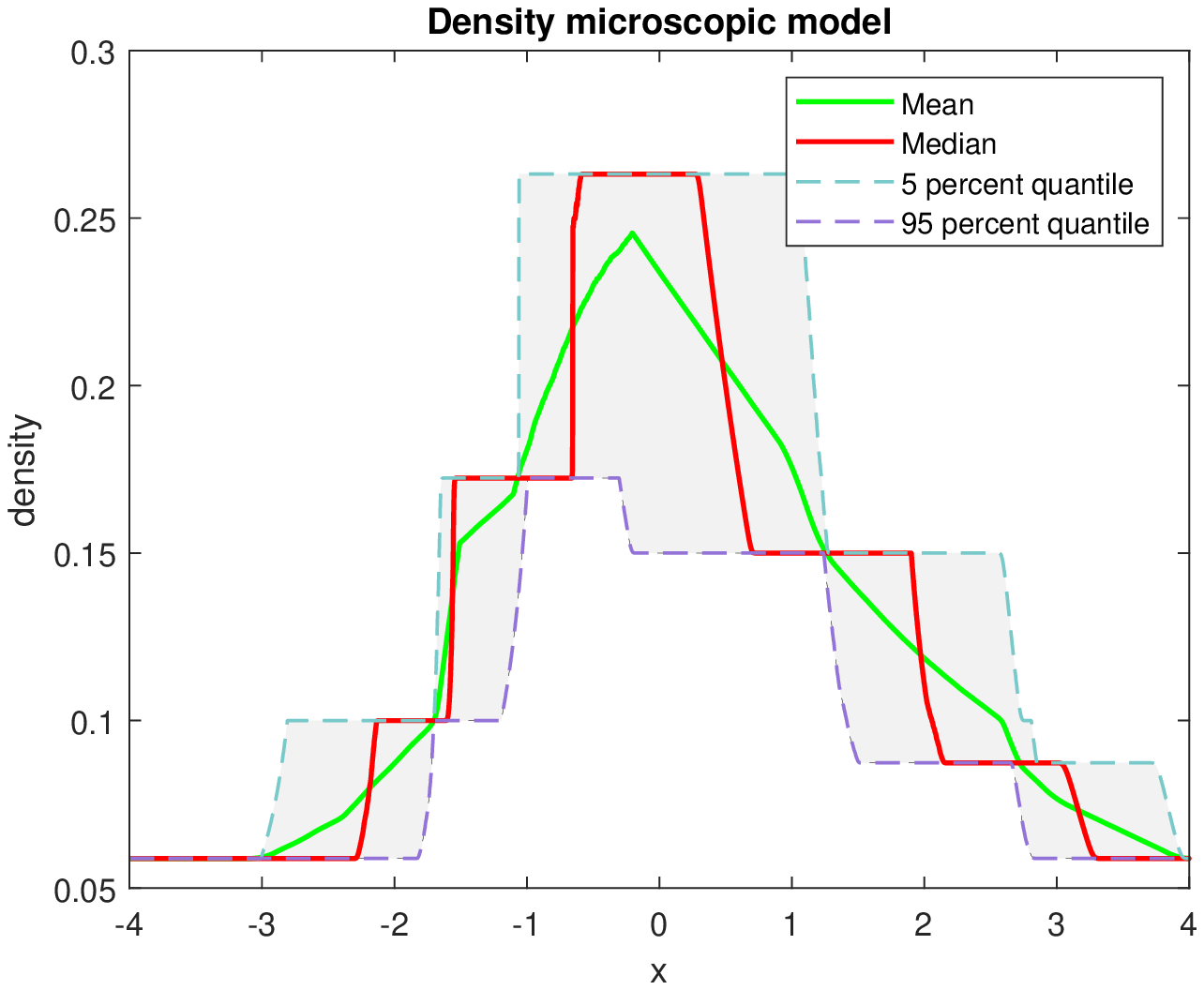}
    \captionof{figure}{Mean, median and 90 percent confidence interval of the density evolution of the microscopic model \eqref{eq:micro_model_Y} at $T=10$ for uniformly distributed $Y$.}
    \label{fig:confidenceMicro}
\end{minipage} \\
\begin{minipage}{0.49\textwidth}
    \centering
    \includegraphics[width=\textwidth]{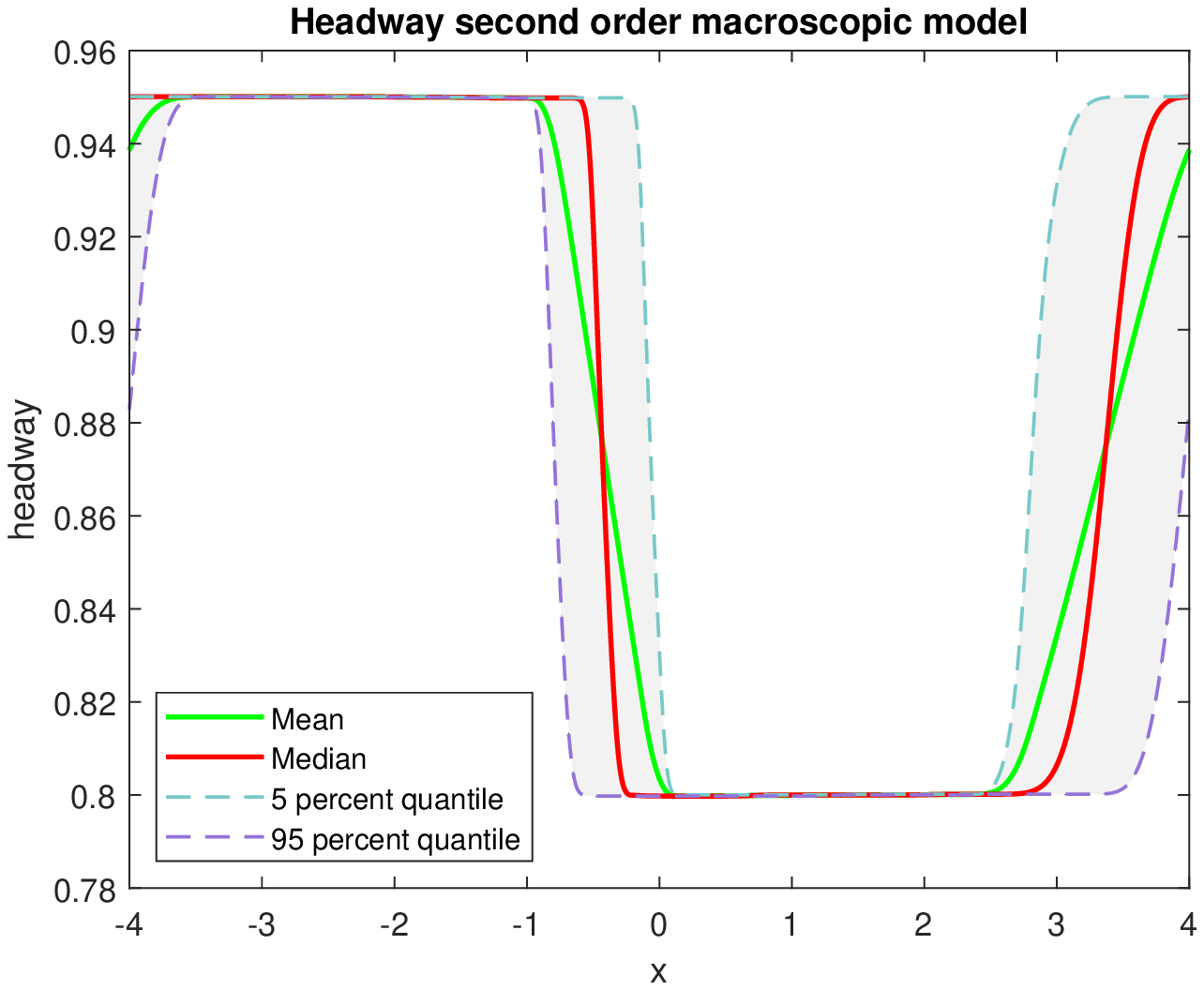}
    \captionof{figure}{Mean, median and 90 percent confidence interval of the headway evolution of the second order macroscopic model \eqref{eq:random_macro2} at $T=10$ for uniformly distributed $Y$.}
    \label{fig:confidenceMacro_headway}
\end{minipage}~
\begin{minipage}{0.49\textwidth}
    \centering
    \includegraphics[width=\textwidth]{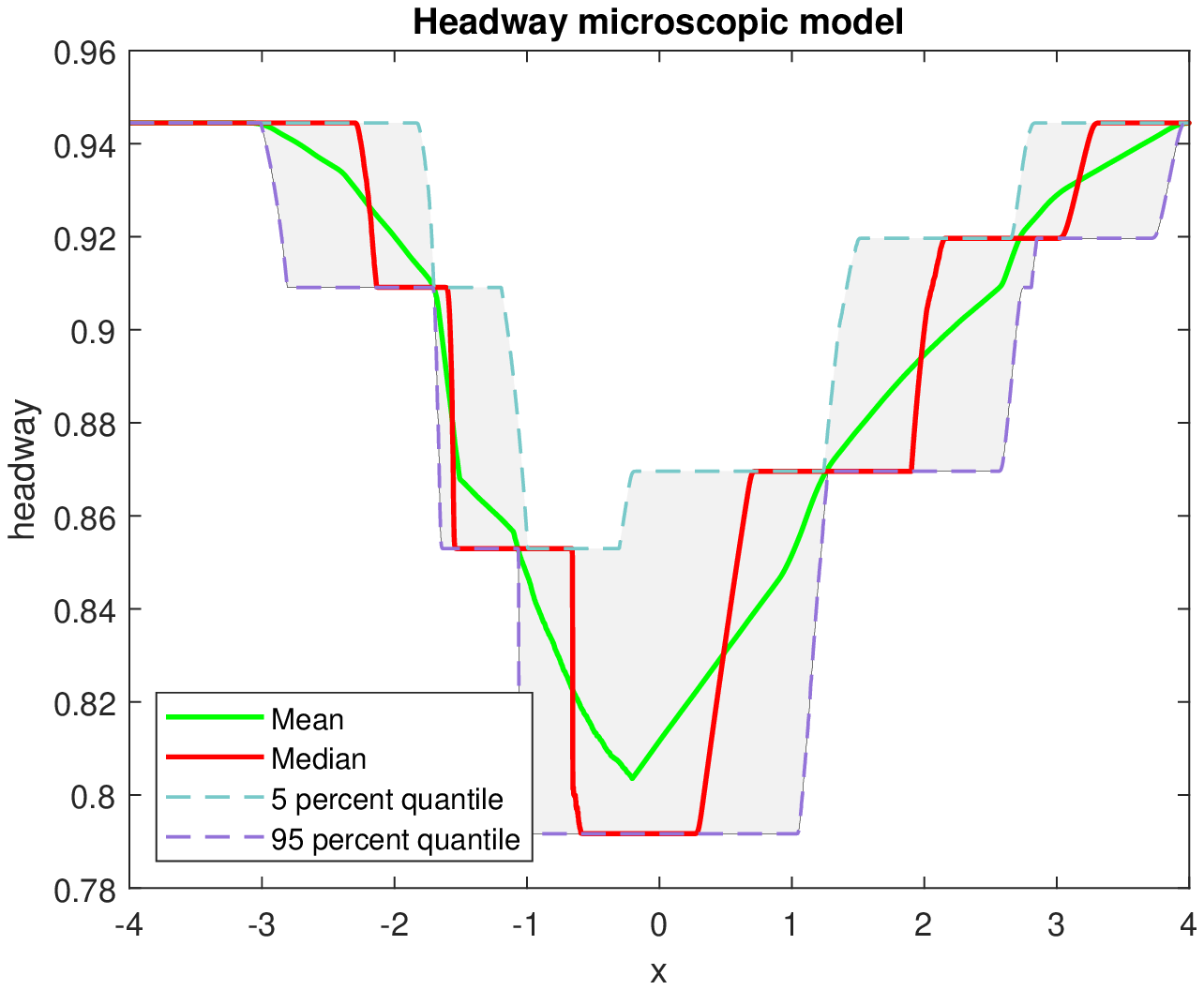}
    \captionof{figure}{Mean, median and 90 percent confidence interval of the headway evolution of the microscopic model \eqref{eq:micro_model_Y} at $T=10$ for uniformly distributed $Y$.}
    \label{fig:confidenceMicro_headway}
\end{minipage}
\end{figure}

To describe the perturbed systems, we consider not only the mean realization of the densities, but confidence intervals in the Figures \ref{fig:confidenceMacro} and \ref{fig:confidenceMicro} {in the special case of the uniform distribution ($\alpha=\beta=1$)}. The upper dashed lines show the level of the five percent highest densities in the Monte Carlo run, whereas the lower dashed lines represent the five percent lowest densities in the simulation. The green curve shows again the mean realization, whereas the red curve shows the median representing the 50 percent highest densities.

We observe that there are areas in which densities and headways vary in a small range and are almost deterministic. But there are also road sections in which the densities show a large variance, as can be seen in Figure \ref{fig:confidenceMacro} for the macroscopic model in the range of $x\in[-1,0]$. Due to the density increases that have already been observed in Figure \ref{fig:comparison_a=0} high densities are attained frequently. This is not the case in the microscopic simulation in Figure \ref{fig:confidenceMicro}. It is also striking that the headways in Figure \ref{fig:confidenceMacro_headway} of the second order macroscopic model behave very stable with regard to the changed accident sizes.

{To shortly illustrate the behaviour for a different distribution of $Y=2Z+1$, we set $\alpha=5, \beta=2$ for $Z$ corresponding to a right-skewed Beta distribution of the accident size on the interval $[1,3]$. The results for the densities of the second order macroscopic model~\eqref{eq:random_macro2} and the microscopic model \eqref{eq:micro_model_Y} are presented in Figures \ref{fig:confidenceMacro_beta} - \ref{fig:confidenceMicro_headway_beta}. In comparison to the uniform distribution, we mainly observe two effects: on the one hand, the increase of the densities is slightly shifted to the left due to the right-skewed distribution of the accident size that makes larger accidents more likely. On the other hand, the 90 percent confidence intervals of the density and headway are thinner due to the probability density function decaying to zero as we reach the boundary values of $Y$. But the overall shape of the quantities is very similar to the results for the uniform distribution.

\begin{figure}[!t]
\begin{minipage}{0.49\textwidth}
    \centering
    \includegraphics[width=\textwidth]{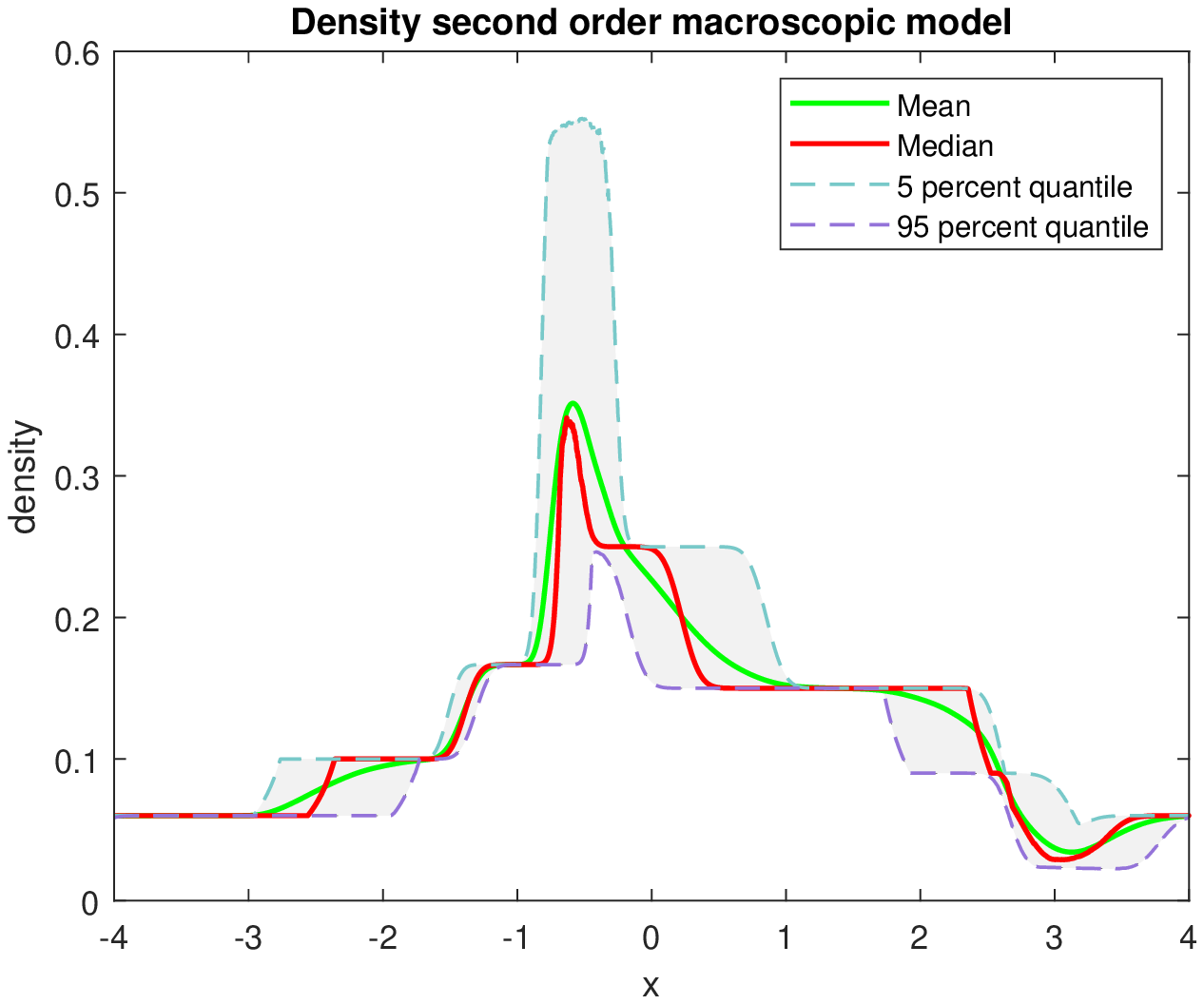}
    \captionof{figure}{Mean, median and 90 percent confidence interval of the density evolution of the second order macroscopic model \eqref{eq:random_macro2} at $T=10$ for shifted and scaled beta distributed $Y$.}
    \label{fig:confidenceMacro_beta}
\end{minipage}~
\begin{minipage}{0.49\textwidth}
    \centering
    \includegraphics[width=\textwidth]{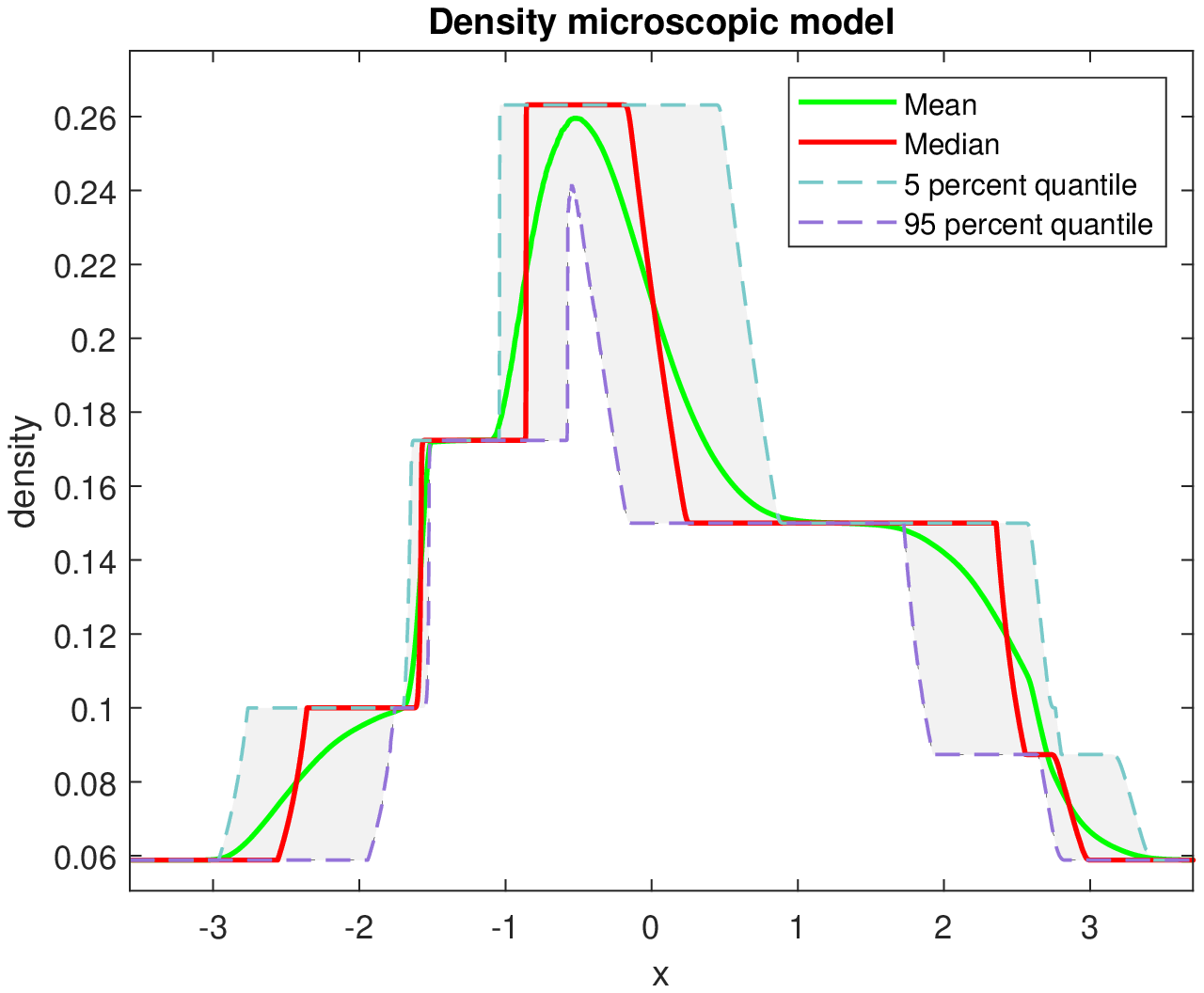}
    \captionof{figure}{Mean, median and 90 percent confidence interval of the density evolution of the microscopic model \eqref{eq:micro_model_Y} at $T=10$ for shifted and scaled beta distributed $Y$.}
    \label{fig:confidenceMicro_beta}
\end{minipage} \\
\begin{minipage}{0.49\textwidth}
    \centering
    \includegraphics[width=\textwidth]{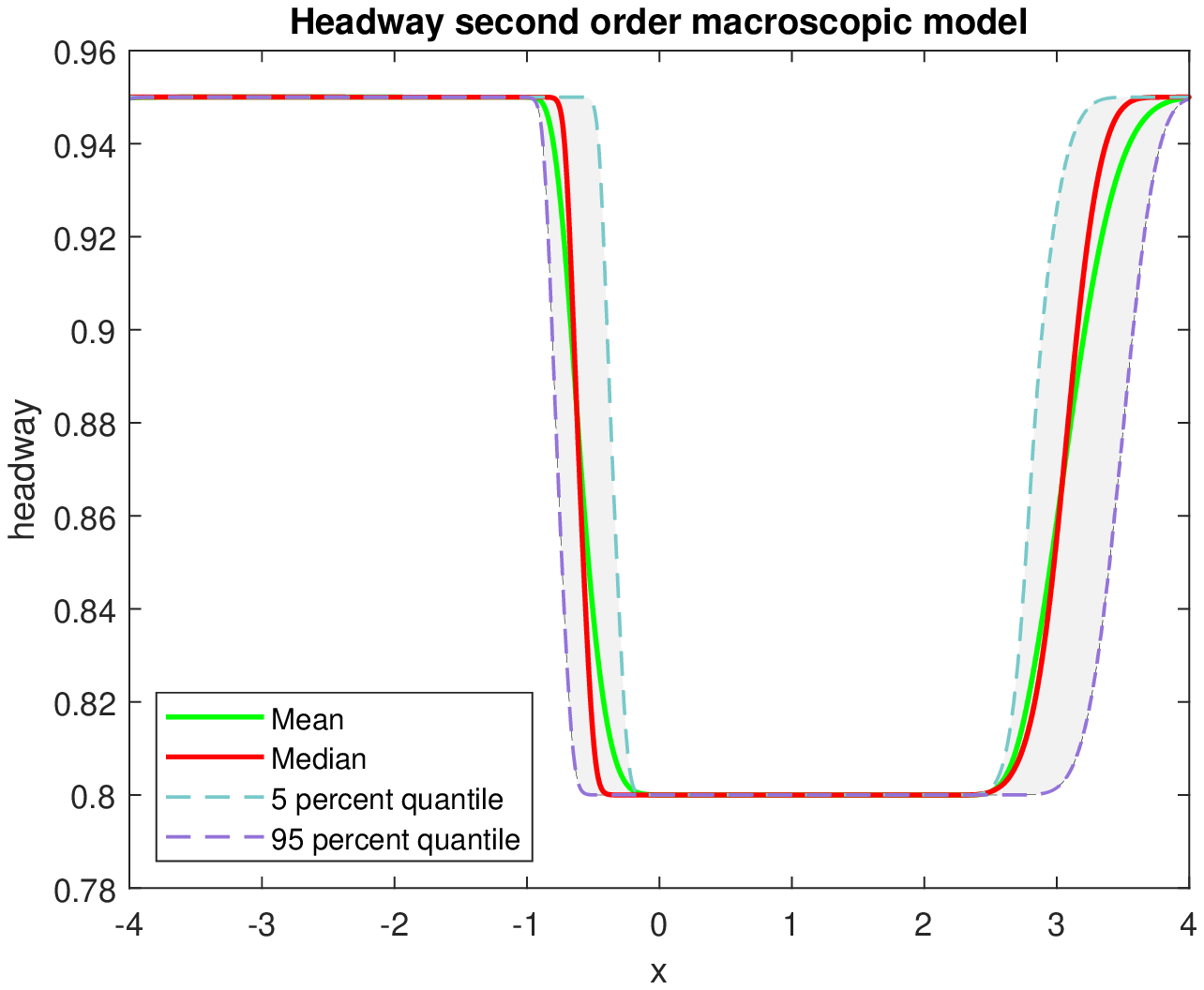}
    \captionof{figure}{Mean, median and 90 percent confidence interval of the headway evolution of the second order macroscopic model \eqref{eq:random_macro2} at $T=10$ for shifted and scaled beta distributed $Y$.}
    \label{fig:confidenceMacro_headway_beta}
\end{minipage}~
\begin{minipage}{0.49\textwidth}
    \centering
    \includegraphics[width=\textwidth]{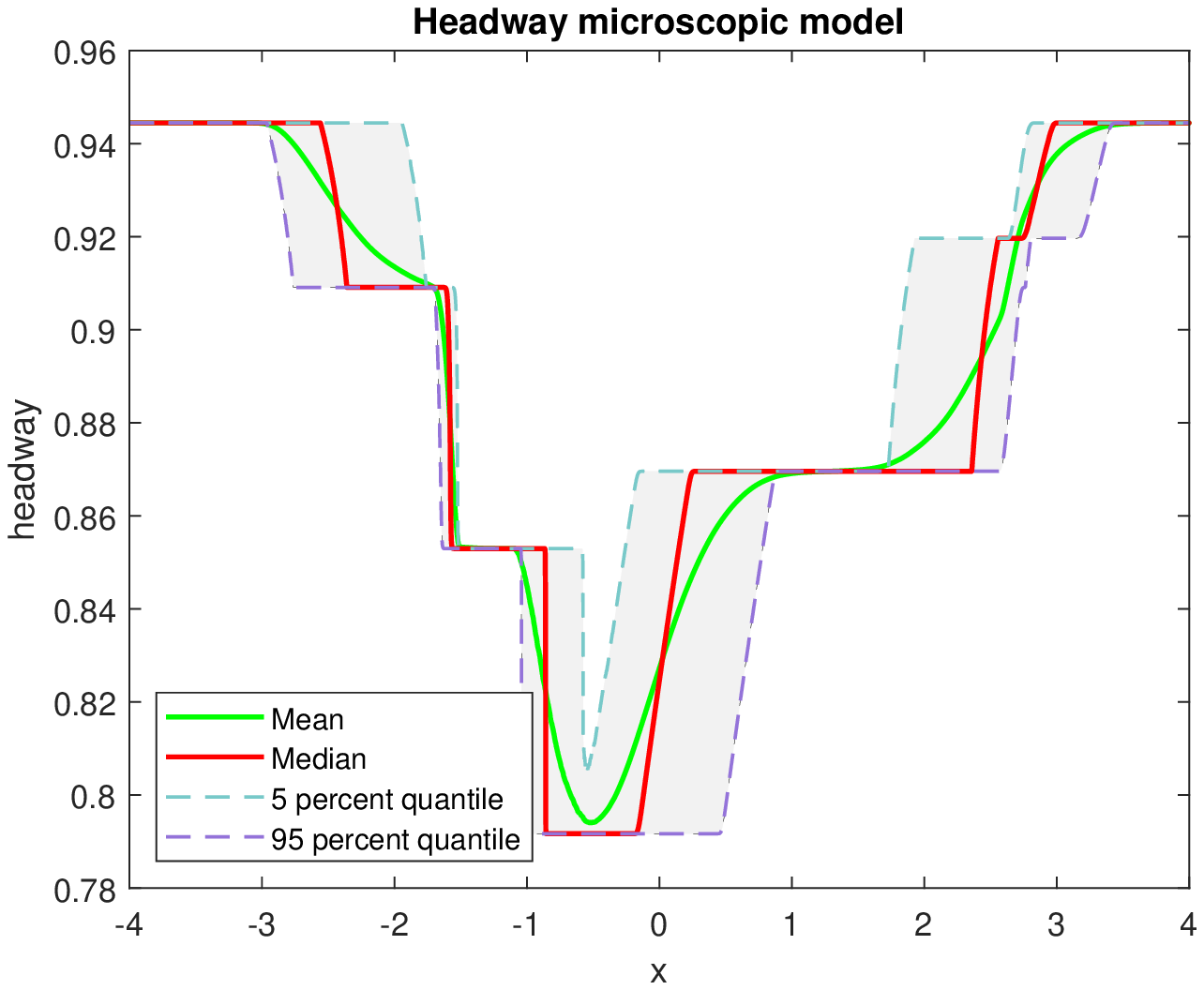}
    \captionof{figure}{Mean, median and 90 percent confidence interval of the headway evolution of the microscopic model \eqref{eq:micro_model_Y} at $T=10$ for shifted and scaled beta distributed $Y$.}
    \label{fig:confidenceMicro_headway_beta}
\end{minipage}
\end{figure}
}
		
{Apart from the Monte Carlo simulations, we approximate the expectations of the densities in both models using a polynomial chaos expansion. Note that we are not particularly interested in providing a deep analysis on the polynomial chaos expansion itself but exploit this approach as an alternative way of approximating the expected value of the densities. 

From now on we stick to the case of $Y$ being uniformly distributed in $[1,3]$. The Legendre polynomials form an orthonormal basis for a uniformly distributed random variable on $[-1,1]$. For our purpose a simple transformation by an additive shift transforms $Y$ to the interval $[1,3]$. Then orthonormality means
\begin{align*}
    \int_{1}^3 \frac{1}{2} \phi_i(y) \phi_j(y) dy = \delta_{i,j},
\end{align*} where $\phi_k$ denotes the $k$-th Legendre polynomial.} 
{ We consider the polynomial chaos expansion for the second order macroscopic model in conservative form \eqref{eq:macroheadway_conservative1}, where 
\begin{align}
\label{eq: defZ}
z(x,t) =\rho(x,t)\left(h(x,t) + \frac{\gamma}{2}\eta \rho(x,t)\right).
\end{align}
Then, the system is described in the solution subspace $\text{span}\left\lbrace\phi_k\right\rbrace_{k=0}^K, K \in \N$ by
\begin{align*}
    \begin{pmatrix}
        \rho^K(x,t;Y)\\z^K(x,t;Y)
    \end{pmatrix} = \sum_{k=0}^K \begin{pmatrix}
        \hat{\rho}_k(x,t) \phi_k(Y)\\ \hat{z}_k(x,t) \phi_k(Y)
    \end{pmatrix},
\end{align*}
where at the initial time $\hat{\rho}_k(x,0)$ and $\hat{z}_k(x,0)$ are the modes of the expansion and can be determined as follows:
\begin{align}
\label{eq:defModes}
    \hat{\rho}_k(x,0) = \int_1^3 \frac{1}{2}\rho(x,0,y)\phi_k(y)dy,~~~ \hat{z}_k(x,0) = \int_1^3 \frac{1}{2}z(x,0,y)\phi_k(y)dy.
\end{align}
The propagation of the modes up to order $K$ can be described by the system
\begin{align}\begin{split}
    \label{eq: GalerikinSystem}
 &0=\partial_t \begin{pmatrix}
     \hat{\rho}_0(x,t)\\ \hat{z}_0(x,t) \\ \vdots \\ \hat{\rho}_K(x,t)\\ \hat{z}_K(x,t)
 \end{pmatrix} \\
 &+ \partial_x \begin{pmatrix}
     \int_1^3  c(x)V\left( \frac{\sum_{k=0}^K\hat{z}_k(x,t)\phi_k(y)}{\sum_{k=0}^K\hat{\rho}_k(x,t)\phi_k(y)} - \frac{\gamma}{2}\eta\sum_{k=0}^K\hat{\rho}_k(x,t)\phi_k(y)\right)(\sum_{k=0}^K\hat{\rho}_k(x,t)\phi_k(y))\phi_0(y) \frac{1}{2}dy\\
     \int_1^3 c(x)V\left( \frac{\sum_{k=0}^K\hat{z}_k(x,t)\phi_k(y)}{\sum_{k=0}^K\hat{\rho}_k(x,t)\phi_k(y)} - \frac{\gamma}{2}\eta\sum_{k=0}^K\hat{\rho}_k(x,t)\phi_k(y)\right)(\sum_{k=0}^K\hat{z}_k(x,t)\phi_k(y))\phi_0(y) \frac{1}{2}dy \\ \vdots \\
     \int_1^3  c(x)V\left( \frac{\sum_{k=0}^K\hat{z}_k(x,t)\phi_k(y)}{\sum_{k=0}^K\hat{\rho}_k(x,t)\phi_k(y)} - \frac{\gamma}{2}\eta\sum_{k=0}^K\hat{\rho}_k(x,t)\phi_k(y)\right)(\sum_{k=0}^K\hat{\rho}_k(x,t)\phi_k(y))\phi_K(y) \frac{1}{2}dy\\
     \int_1^3  c(x)V\left( \frac{\sum_{k=0}^K\hat{z}_k(x,t)\phi_k(y)}{\sum_{k=0}^K\hat{\rho}_k(x,t)\phi_k(y)} - \frac{\gamma}{2}\eta\sum_{k=0}^K\hat{\rho}_k(x,t)\phi_k(y)\right)(\sum_{k=0}^K\hat{z}_k(x,t)\phi_k(y))\phi_K(y) \frac{1}{2}dy
 \end{pmatrix}.
\end{split}
\end{align} 
Similarly, we can set up the system for the expansion in the case of the microscopic model \eqref{eq:micro_model_Y} by
%\begin{align*}
%        \rho^{(N),K}(x,t;Y) = \sum_{k=0}^K 
%        \hat{\rho}_k^{(N)}(x,t) \phi_k(Y),
%\end{align*}
\begin{align*}
    x_i^{(N)}(t,Y) = \sum_{k=0}^K \hat{x}_{i,k}^{(N)}(t)\Phi_k(Y),
\end{align*}
where the modes for $k=0,\dots,K$ at the initial time $t=0$ are given by
\begin{align}
\label{eq:defModesMic}
    %\hat{\rho}_k^{(N)}(x,0) = \int_1^3 \rho^{(N)}(x,0,y)\phi_k(y)\frac{1}{2}dy.
    \hat{x}_{i,k}^{(N)}(0) = \int_1^3 \frac{1}{2}x_i^{(N)}(0)\Phi_k(y)dy.
\end{align}
Then, the evolution of the microscopic system is for $r=0\dots,K$ given by
\begin{align}
\begin{split}
        &\int_1^3 \frac{1}{2} \sum_{k=0}^K  \dot{\hat{x}}_{i,k}^{(N)}(t)\Phi_k(y)\Phi_r(y) dy \\
    &= \int_1^3 \frac{1}{2} c\left(\sum_{k=0}^K \hat{x}_{i,k}^{(N)}(t)\Phi_k(y)\right) \tilde{V}\left( \frac{L}{\sum_{k=0}^K\left(\hat{x}^{(N)}_{i+1,k}(t) - \hat{x}^{(N)}_{i,k}(t)\right)\Phi_k(y)}  \right)\Phi_r(y)dy,
    \label{eq: GalerkinMicro1}
\end{split}
\end{align}
which can by orthogonality of the basis functions for $r=0\dots,K$ be rewritten to 
\begin{align}
\begin{split}
\label{eq: GalerkinMicro2}
    & \dot{\hat{x}}_{i,r}^{(N)}(t)= \int_1^3 \frac{1}{2} c\left(\sum_{k=0}^K \hat{x}_{i,k}^{(N)}(t)\Phi_k(y)\right) \tilde{V}\left( \frac{L}{\sum_{k=0}^K\left(\hat{x}^{(N)}_{i+1,k}(t) - \hat{x}^{(N)}_{i,k}(t)\right)\Phi_k(y)}  \right)\Phi_r(y)dy.
\end{split}
\end{align}

The vehicle positions can be transformed to the microscopic local density function by the equations \eqref{eq: locDensities} and \eqref{eq: locDensities2}.

 As we aim to approximate the expectation of the densities, we focus on the modes for $K=0$ which exactly describe the expectation of the stochastic system. The integrals in \eqref{eq:defModes}-\eqref{eq: GalerkinMicro2} can be approximated using Gauss-Legendre quadrature.}
%As we want to approximate the expectation of the densities, which is an integral of the form~\eqref{eq:expectationMacro} with respect to the uniform probability density $g(y)=\frac{1}{2}\mathbb{1}_{[1,\,3]}(y)$, using Gauss-Legendre quadrature we can perform the analysis only for the roots of the $n$-th Legendre polynomial (shifted additively by 2) which reduces the computational effort compared to a Monte Carlo estimation drastically. 
The weights for the corresponding quadrature rule and a root $z_k$ are given by (staying on the roots inside $[-1,1]$)
$$ w_k=\frac{1}{(1-z_k^2)\cdot(L_n^\prime (z_k))^2}, $$
where $L_n^\prime$ is the derivative of the $n$-th Legendre polynomial. The expectation of the macroscopic density at a position $x$ and time $t$ given by %in \eqref{eq:truncExpectation}
\begin{equation*}
   % \label{eq:truncExpectation}
    \mathbb{E}_Y[\rho(x,T;Y)] {=\hat{\rho}_0(x,T)} .
\end{equation*}
Similarly, an approximation of the expected headway is given by
\begin{equation*}
   % \label{eq:truncExpectationHeadway}
    \mathbb{E}_Y[h(x,T;Y)] {=\hat{h}_0(x,T)},
\end{equation*}
where $\hat{h}_0$ is recovered from $\hat{z}_0$ by \eqref{eq: defZ}. 
One can proceed exactly in the same way for the microscopic model and approximate
$$ \mathbb{E}_Y[\rho^{(N)}(x,T;Y)] { =\hat{\rho}_0^{(N)}(x,T)}, $$
where $\hat{\rho}_0^{(N)}(x,T)$ is the piecewise constant density function resulting from the vehicle position modes $\hat{x}_{i,0}^{(N)}(T)$ using \eqref{eq: locDensities} and \eqref{eq: locDensities2}. 
The expectation of the headway in the microscopic model is obtained when plugging the density into the optimal headway function from \eqref{eq:def_velo_headway}.

\begin{figure}[!t]
\begin{minipage}{0.49\textwidth}
    \includegraphics[width=\textwidth]{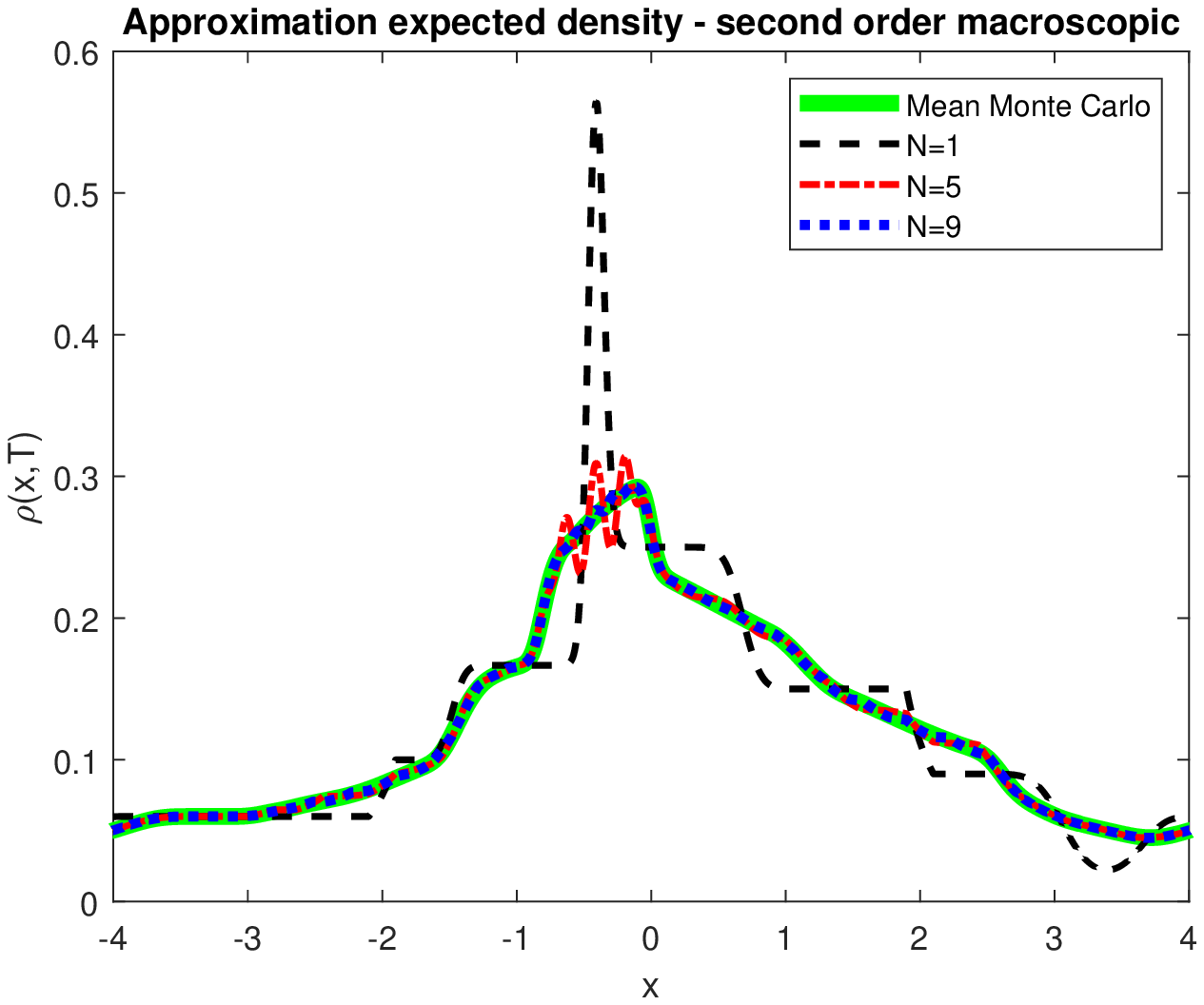}
    \captionof{figure}{Approximation of the expected densities in the second order macroscopic model \eqref{eq:random_macro2} using polynomial chaos expansion.}
    \label{fig:DiscretMacro}
\end{minipage}~
\begin{minipage}{0.49\textwidth}
    \includegraphics[width=\textwidth]{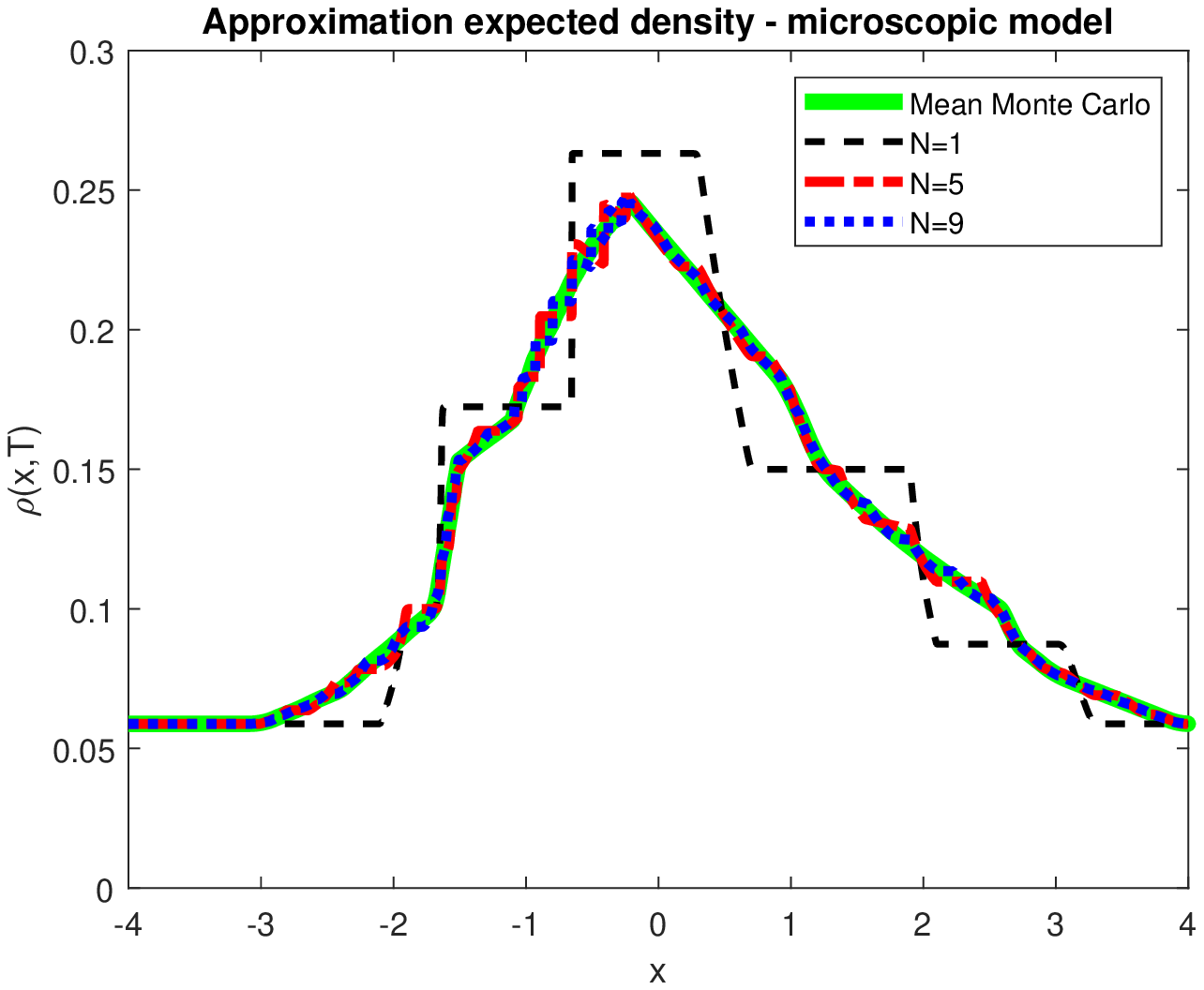}
    \captionof{figure}{Approximation of the expected densities in the microscopic model \eqref{eq:micro_model_Y} using polynomial chaos expansion.}
    \label{fig:DiscretMicro}
\end{minipage} \\
\begin{minipage}{0.49\textwidth}
    \includegraphics[width=\textwidth]{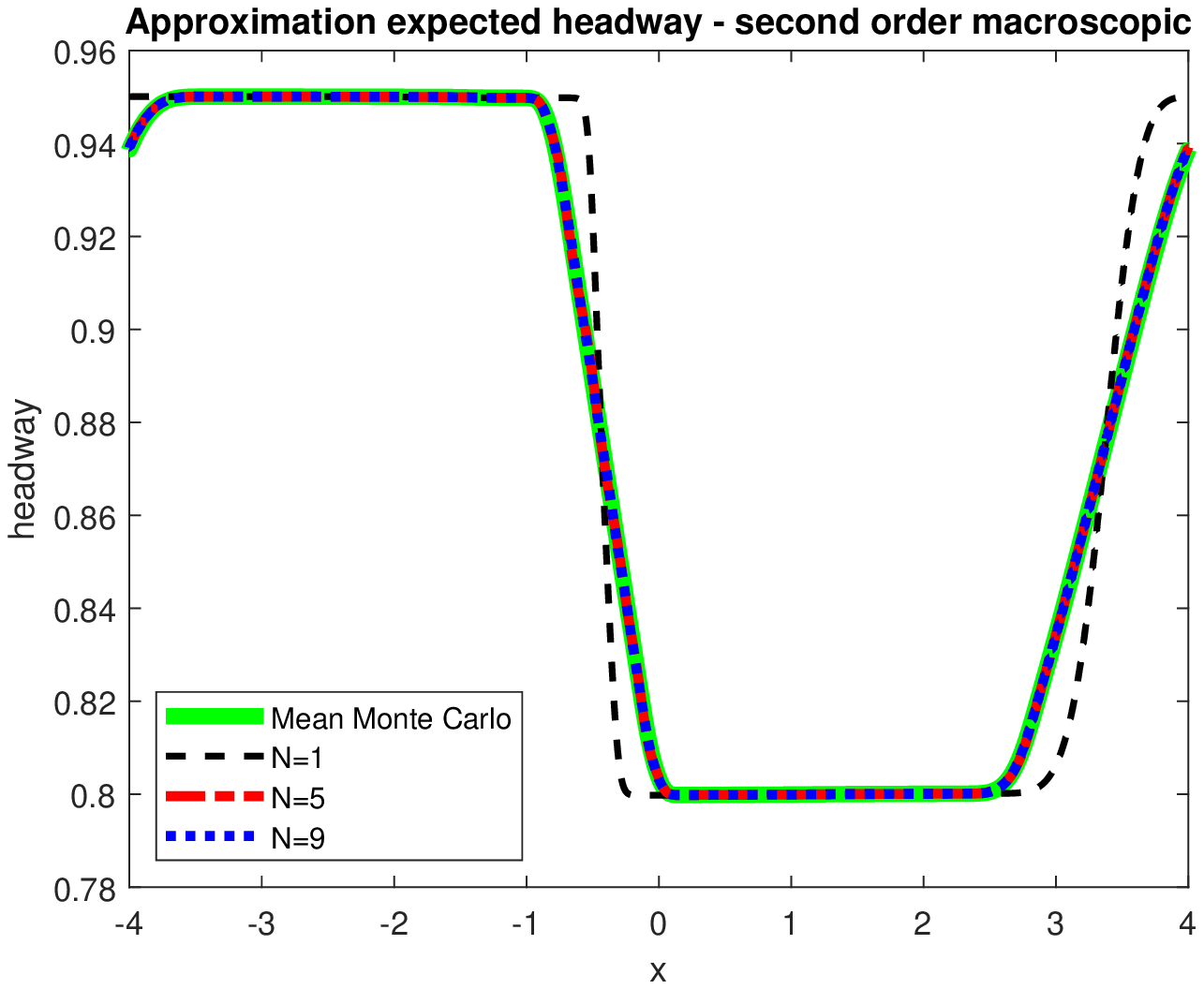}
    \captionof{figure}{Approximation of the expected headways in the second order macroscopic model \eqref{eq:random_macro2} using polynomial chaos expansion.}
    \label{fig:DiscretMacro_headway}
\end{minipage}~
\begin{minipage}{0.49\textwidth}
    \includegraphics[width=\textwidth]{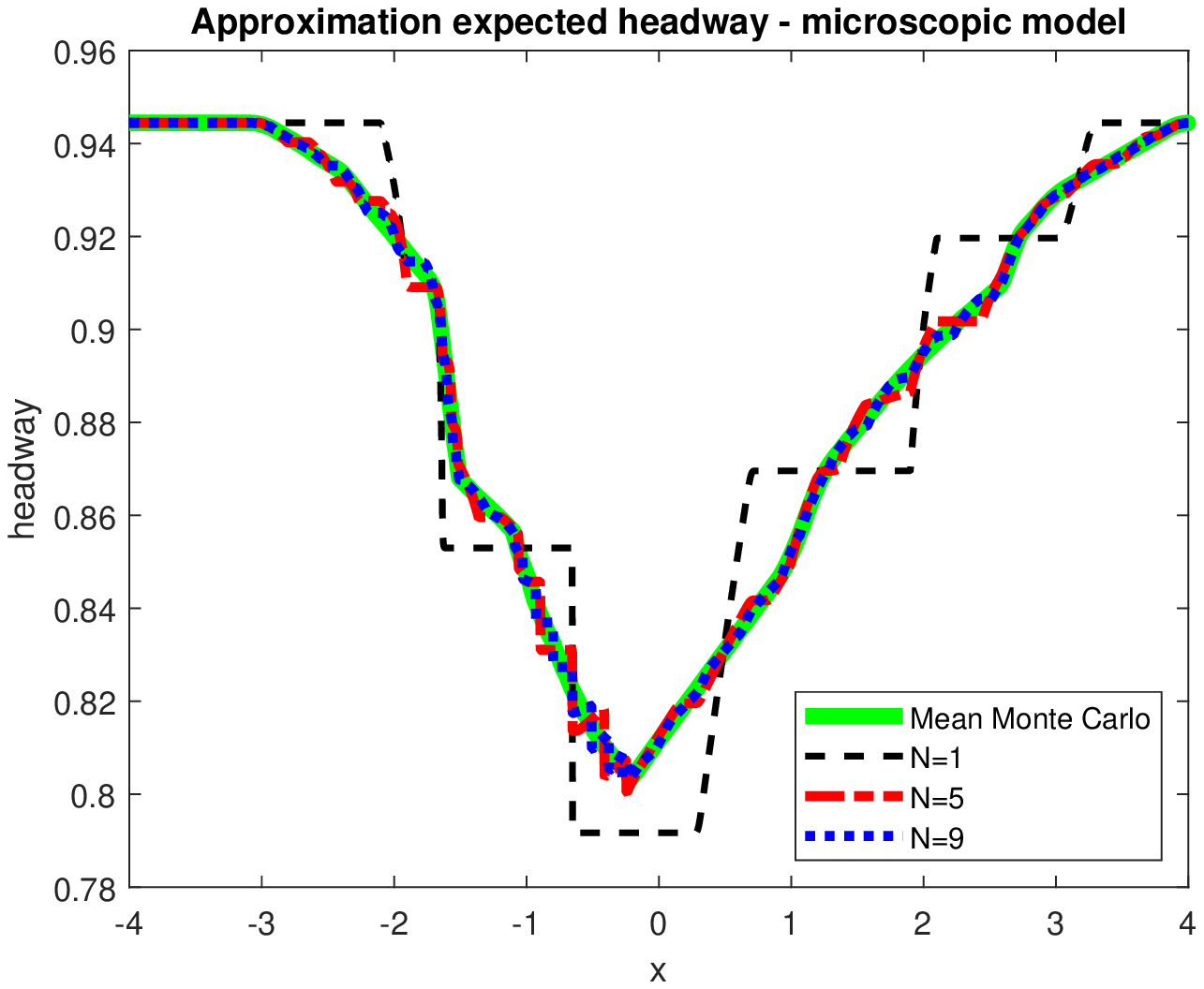}
    \captionof{figure}{Approximation of the expected headways in the microscopic model \eqref{eq:micro_model_Y} using polynomial chaos expansion.}
    \label{fig:DiscretMicro_headway}
\end{minipage}
\end{figure}
  
In the Figures \ref{fig:DiscretMacro} and \ref{fig:DiscretMicro} we show the approximations for the expectations using $n=1,5,9$ roots of the Legendre polynomials and compare it with the expectation of the Monte Carlo simulations with $2\cdot 10^3$ realizations. For $n=1$ the approximation coincides with the one choosing the accident size $Y$ to $\mathbb{E}[Y]=2$ and is not very accurate on both levels. Overall for $n=5$ the approximation shows a good performance but has still some inaccuracies in some areas, as for example in the macroscopic density around $x=-0.5$. For $n=9$ the approximation using the polynomial chaos expansion shows no discernible differences for both the macroscopic and microscopic expected density.

\begin{figure}[!t]
\begin{minipage}{0.49\textwidth}
    \includegraphics[width=\textwidth]{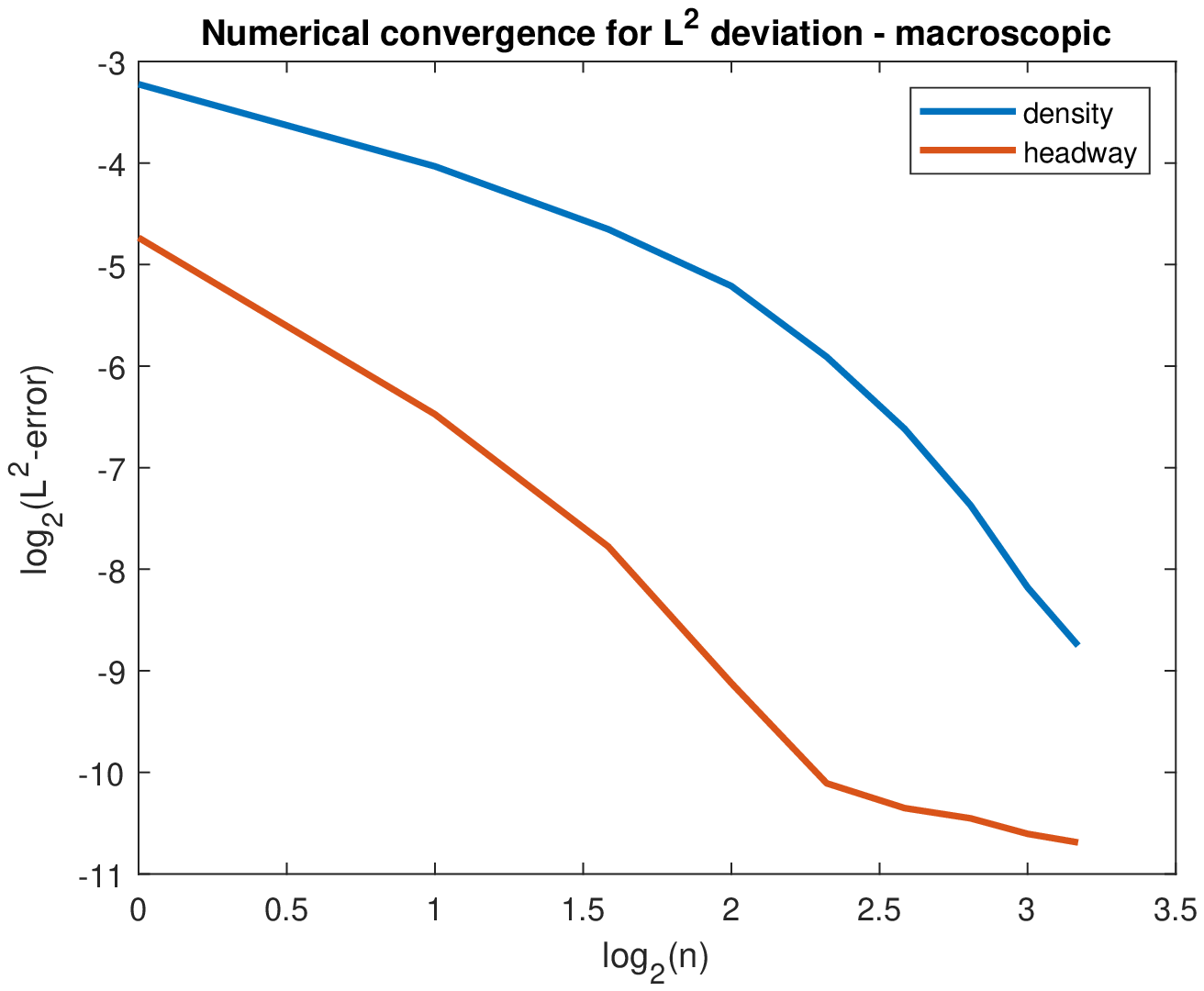}
    \captionof{figure}{Evolution of the $L^2$-error in density and headway using the generalized polynomial chaos expansion for the second order macroscopic model \eqref{eq:random_macro2} and $n=1,\dots,9$ nodes.}
    \label{fig:ConvergenceMacro}
\end{minipage}~
\begin{minipage}{0.49\textwidth}
    \includegraphics[width=\textwidth]{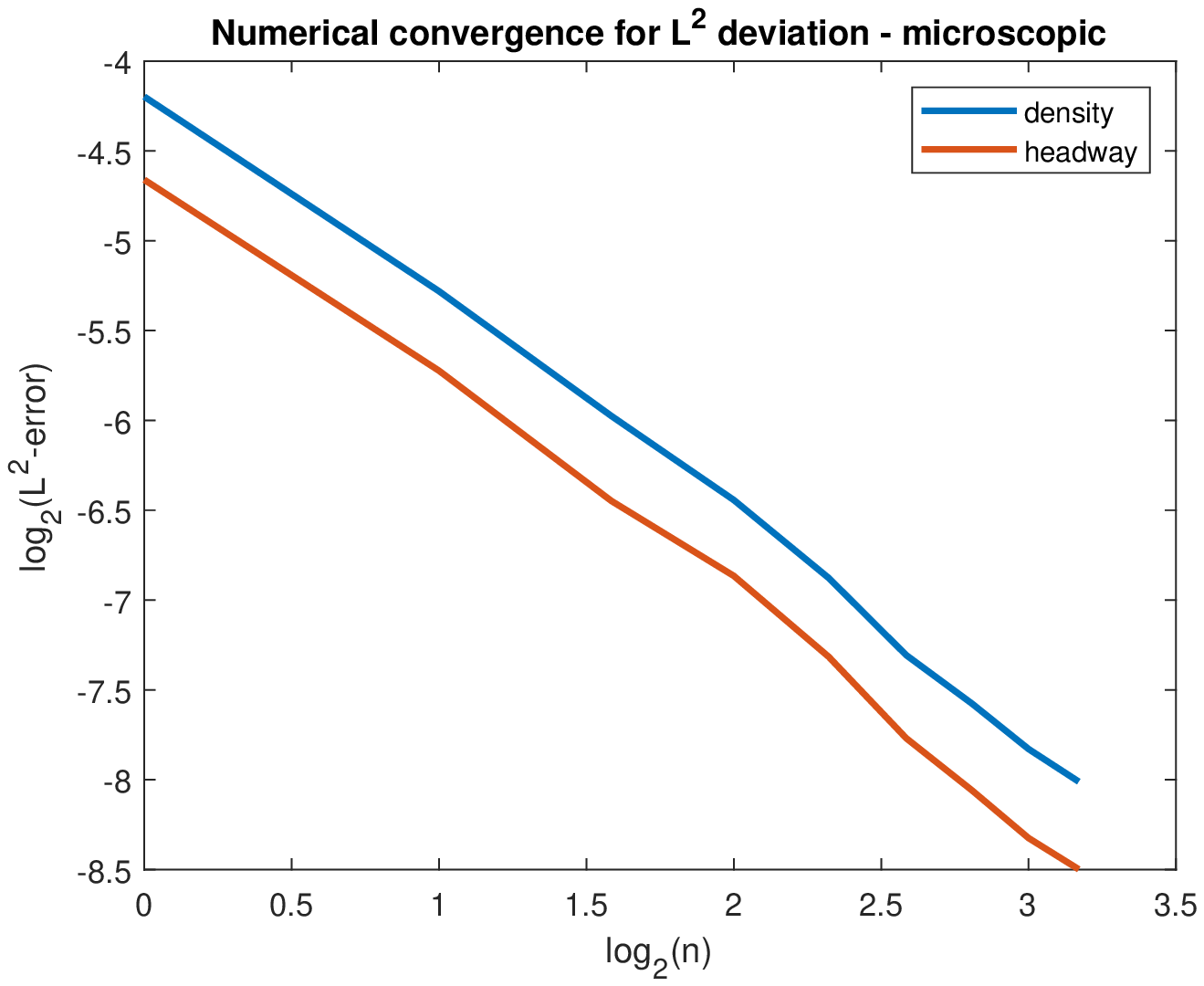}
    \captionof{figure}{Evolution of the $L^2$-error in density and headway using the generalized polynomial chaos expansion for the microscopic model \eqref{eq:micro_model_Y} and $n=1,\dots,9$ nodes.}
    \label{fig:ConvergenceMicro}
\end{minipage}
\end{figure}

A convergence analysis of the expectation approximated using $n=1,\dots,9$ nodes in the polynomial chaos approach towards the result from the Monte Carlo simulation in a framework with logarithmic values on the axis is presented in the Figures \ref{fig:ConvergenceMacro} and \ref{fig:ConvergenceMicro}. On the microscopic scale we recognize a linear relationship with approximate convergence rates of slightly larger than 2, whereas the correlation on the macroscopic scale is not perfectly linear but still strictly decreasing with an approximate convergence rate of 2. These results underline the approximation behaviour in the Figures \ref{fig:confidenceMacro}-\ref{fig:confidenceMicro_headway}.

\section{Conclusions}
\label{sect:Conclusion}
In this paper, we have considered classical Follow-the-Leader traffic dynamics with space-dependent speed, which, upon reformulation as interaction dynamics of a system of stochastic particles, we have described at the mesoscopic scale by means of an Enskog-type collisional kinetic equation. Since Follow-the-Leader dynamics are based on the reciprocal distance of the vehicles, in our kinetic representation we have used the \textit{headway}, along with the position, as microscopic state of the vehicles in place of the more usual speed. This has allowed us to obtain formally, in the hydrodynamic limit, macroscopic conservation laws based on the density and the \textit{mean headway} of traffic, which constitute original models with respect to those consolidated in the reference literature. In particular, our investigations have shown that, in the limit, one may formally get either a first or a second order macroscopic model with space-dependent flux depending on a certain relaxation parameter of the headway. Analytical investigations have proved that our new second order model complies with the Aw-Rascle consistency condition and admits weak entropy solutions at least for initial data with small total variation.

We have used these traffic models to describe the impact of accidents on the traffic flow. Our derivation from principles of statistical mechanics has made it possible to include, in particular, accidents taking place in \textit{random}, viz. \textit{uncertain}, positions along the road. In the hydrodynamic limit, this has yielded a new version of the former second order macroscopic model with \textit{uncertain flux}. Numerical investigations on uncertain accident sizes have illustrated, on one hand, that expected traffic densities can be computed efficiently using a polynomial chaos expansion and, on the other hand, that some road sections may be much more affected by accidents, hence may face a much larger variety of traffic scenarios, than others.

Future work may consider generalised space-dependent traffic accident models on road networks with \textit{ad-hoc} numerical simulation techniques.

\section*{Acknowledgments}
F.A.C. is member of Gruppo Nazionale per l'Analisi Matematica, la Probabilit\`a e le loro Applicazioni (GNAMPA) of the Istituto Nazionale di Alta Matematica (INdAM). F.A.C. was partially supported by the Ministry of University and Research (MUR), Italy, under the grant PRIN 2020 - Project N. 20204NT8W4, ``Nonlinear evolution PDEs, fluid dynamics and transport equations: theoretical foundations and applications''. F.A.C. would like to thank Debora Amadori for useful discussions about the analytical properties. S. G. was supported by the German Research Foundation (DFG) under grant GO 1920/10-1, 11-1 and 12-1. A.T. is member of Gruppo Nazionale per la Fisica Matematica (GNFM) of INdAM, Italy.

\bibliographystyle{plain}
%\bibliography{Accidents}

\begin{thebibliography}{10}
	
	\bibitem{aw2000SIAP}
	A.~Aw and M.~Rascle.
	\newblock Resurrection of ``second order'' models of traffic flow.
	\newblock {\em SIAM J. Appl. Math.}, 60(3):916--938, 2000.
	
	\bibitem{Brencher2020}
	L.~Brencher and A.~Barth.
	\newblock {H}yperbolic {C}onservation {L}aws with {S}tochastic {D}iscontinuous
	{F}lux {F}unctions.
	\newblock In {\em International Conference on Finite Volumes for Complex
		Applications}, pages 265--273. Springer, 2020.
	
	\bibitem{Bressan}
	A.~Bressan.
	\newblock {\em Hyperbolic Systems of Conservation Laws. The one-dimensional
		Cauchy problem.}, volume~20.
	\newblock Oxford Lecture Series in Mathematics and its Applications. Oxford
	University Press, 2000.
	
	\bibitem{chiarello2021MMS}
	F.~A. Chiarello, B.~Piccoli, and A.~Tosin.
	\newblock Multiscale control of generic second order traffic models by
	driver-assist vehicles.
	\newblock {\em Multiscale Model. Simul.}, 19(2):589--611, 2021.
	
	\bibitem{MMS2021}
	F.~A. Chiarello, B.~Piccoli, and A.~Tosin.
	\newblock Multiscale control of generic second order traffic models by
	driver-assist vehicles.
	\newblock {\em Multiscale Model. Simul.}, 19(2):589--611, 2021.
	
	\bibitem{chiarello2023KRM}
	F.~A. Chiarello and A.~Tosin.
	\newblock Macroscopic limits of non-local kinetic descriptions of vehicular
	traffic.
	\newblock {\em Kinet. Relat. Models}, 16(4):540--564, 2023.
	
	\bibitem{DafermosGeng}
	C.~M. Dafermos and X.~Geng.
	\newblock Generalized characteristics uniqueness and regularity of solutions in
	a hyperbolic system of conservation laws.
	\newblock {\em Annales de l'I.H.P. Analyse non linéaire}, 8(3-4):231--269,
	1991.
	
	\bibitem{dimarco2022JSP}
	G.~Dimarco, A.~Tosin, and M.~Zanella.
	\newblock Kinetic derivation of {A}w--{R}ascle--{Z}hang-type traffic models
	with driver-assist vehicles.
	\newblock {\em J. Stat. Phys.}, 186(1):17/1--26, 2022.
	
	\bibitem{freguglia2017}
	P.~Freguglia and A.~Tosin.
	\newblock Proposal of a risk model for vehicular traffic: A {B}oltz-mann-type
	kinetic approach.
	\newblock {\em Commun. Math. Sci.}, 15:213 --236, 2017.
	
	\bibitem{garcia2018}
	B.~Garc\'{\i}a~de Soto, A.~Bumbacher, M.~Deublein, and B.~Adey.
	\newblock Predicting road traffic accidents using artificial neural network
	models.
	\newblock {\em Infrastr. Asset Manag.}, 5(4):132--144, 2018.
	
	\bibitem{gazis1961OR}
	D.~C. Gazis, R.~Herman, and R.~W. Rothery.
	\newblock Nonlinear follow-the-leader models of traffic flow.
	\newblock {\em Oper. Res.}, 9:545--567, 1961.
	
	\bibitem{knapp2020}
	S.~G\"ottlich and S.~Knapp.
	\newblock Modeling random traffic accidents by conservation laws.
	\newblock {\em Math. Biosci. Eng.}, 17:1677--1701, 2020.
	
	\bibitem{ThomasSimone}
	S.~G\"{o}ttlich and T.~Schillinger.
	\newblock Microscopic and macroscopic traffic flow models including random
	accidents.
	\newblock {\em Commun. Math. Sci.}, 19(6):1579--1609, 2021.
	
	\bibitem{herty2018SIAP}
	M.~Herty, A.~Tosin, G.~Visconti, and M.~Zanella.
	\newblock Hybrid stochastic kinetic description of two-dimensional traffic
	dynamics.
	\newblock {\em SIAM J. Appl. Math.}, 78(5):2737--2762, 2018.
	
	\bibitem{herty2021SEMA-SIMAI}
	M.~Herty, A.~Tosin, G.~Visconti, and M.~Zanella.
	\newblock Reconstruction of traffic speed distributions from kinetic models
	with uncertainties.
	\newblock In G.~Puppo and A.~Tosin, editors, {\em Mathematical Descriptions of
		Traffic Flow: Micro, Macro and Kinetic Models}, volume~12 of {\em ICIAM 2019
		SEMA SIMAI Springer Series}, pages 1--16. Springer, 2021.
	
	\bibitem{holden2018}
	H.~Holden and H.~Risebro.
	\newblock Follow-the-leader models can be viewed as a numerical approximation
	to the lighthill-whitham-richards model for traffic flow.
	\newblock {\em Netw. Heterog. Media}, 13:409--421, 2018.
	
	\bibitem{karlsen2004}
	K.~H. Karlsen and J.~D. Towers.
	\newblock Convergence of the lax-friedrichs scheme and stability for
	conservation laws with a discontinuous space-time dependent flux.
	\newblock {\em Chinese Ann. Math.}, 25(03):287--318, 2004.
	
	\bibitem{klar1997JSP}
	A.~Klar and R.~Wegener.
	\newblock Enskog-like kinetic models for vehicular traffic.
	\newblock {\em J. Stat. Phys.}, 87(1-2):91--114, 1997.
	
	\bibitem{lighthill1955PRSLA}
	M.~J. Lighthill and G.~B. Whitham.
	\newblock On kinematic waves. {II}. {A} theory of traffic flow on long crowded
	roads.
	\newblock {\em Proc. R. Soc. Lond. A}, 229(1178):317--345, 1955.
	
	\bibitem{Mora2017}
	E.~Mora~Villaz\'{a}n.
	\newblock {\em A Bayesian Network Approach for Probabilistic Safety Analysis of
		Traffic networks}.
	\newblock PhD thesis, Universidad de Cantabria, 2017.
	
	\bibitem{pareschi2013BOOK}
	L.~Pareschi and G.~Toscani.
	\newblock {\em Interacting {M}ultiagent {S}ystems: {K}inetic equations and
		{M}onte {C}arlo methods}.
	\newblock Oxford University Press, 2013.
	
	\bibitem{prigogine1960OR}
	I.~Prigogine and F.~C. Andrews.
	\newblock A {B}oltzmann-like approach for traffic flow.
	\newblock {\em Operations Res.}, 8(6):789--797, 1960.
	
	\bibitem{prigogine1971BOOK}
	I.~Prigogine and R.~Herman.
	\newblock {\em Kinetic theory of vehicular traffic}.
	\newblock American Elsevier Publishing Co., New York, 1971.
	
	\bibitem{richards1956OR}
	P.~I. Richards.
	\newblock Shock waves on the highway.
	\newblock {\em Operations Res.}, 4:42--51, 1956.
	
	\bibitem{Temple}
	B.~Temple.
	\newblock Systems of conservation laws with invariant submanifolds.
	\newblock {\em Trans. Am. Math. Soc.}, 280(2):781--795, 1983.
	
	\bibitem{tosin2021MCRF}
	A.~Tosin and M.~Zanella.
	\newblock Uncertainty damping in kinetic traffic models by driver-assist
	controls.
	\newblock {\em Math. Control Relat. Fields}, 11(3):681--713, 2021.
	
	\bibitem{zhang2002TRB}
	H.~M. Zhang.
	\newblock A non-equilibrium traffic model devoid of gas-like behavior.
	\newblock {\em Transportation Res. Part B}, 36(3):275--290, 2002.
	
	\bibitem{zhang2003}
	P.~Zhang and R.-X. Liu.
	\newblock Hyperbolic conservation laws with space-dependent flux: I.
	characteristics theory and riemann problem.
	\newblock {\em J. Comput. Appl. Math.}, 156:1--21, 07 2003.
	
	\bibitem{Zhao2019}
	H.~Zhao, H.~Cheng, T.~Mao, and C.~He.
	\newblock Research on traffic accident prediction model based on convolutional
	neural networks in vanet.
	\newblock In {\em 2019 2nd International Conference on Artificial Intelligence
		and Big Data (ICAIBD)}, pages 79--84, 2019.
	
	\bibitem{Zou2017}
	X.~Zou and W.~Yue.
	\newblock A bayesian network approach to causation analysis of road accidents
	using netica.
	\newblock {\em J. Adv. Transp.}, 2017:1--18, 2017.
	
\end{thebibliography}

\end{document}